%% file: main.tex
\documentclass{aastex631}

\usepackage{xcolor}
\usepackage{hyperref}
\usepackage{booktabs}
\usepackage{subfigure}

\newcommand\blast{{\em BLAST}}
\newcommand\blastp{{\em BLASTPol}}

\begin{document}

\title{The BLAST Observatory: A Sensitivity Study for Far-IR Balloon-borne Polarimeters}

\input{authors}



\begin{abstract}

Sensitive wide-field observations of polarized thermal emission from interstellar dust grains will allow astronomers to address key outstanding questions about the life cycle of matter and energy driving the formation of stars and the evolution of galaxies.  Stratospheric balloon-borne telescopes can map this polarized emission at far-infrared wavelengths near the peak of the dust thermal spectrum - wavelengths that are inaccessible from the ground.  In this paper we address the sensitivity achievable by a Super Pressure Balloon (SPB) polarimetry mission, using as an example the Balloon-borne Large Aperture Submillimeter Telescope (BLAST) Observatory. By launching from Wanaka, New Zealand, BLAST Observatory can obtain a 30-day flight with excellent sky coverage - overcoming limitations of past experiments that suffered from short flight duration and/or launch sites with poor coverage of nearby star-forming regions. This proposed polarimetry mission will map large regions of the sky at sub-arcminute resolution, with simultaneous observations at 175, 250, and 350 $\mu m$, using a total of 8274 microwave kinetic inductance detectors.  Here, we describe the scientific motivation for the BLAST Observatory, the proposed implementation, and the forecasting methods used to predict its sensitivity.  We also compare our forecasted experiment sensitivity with other facilities.

%%\footnote{Abstracts for Publications of the Astronomical Society of the Pacific are limited to 300 words}.

%%Measurements of linear polarization at far-IR and submillimeter wavelengths can provide crucial information on interstellar magnetic fields as well as constrainThe Balloon-borne Large Aperture Submillimeter Telescope (BLAST) Observatory is a proposed mission to address key outstanding questions about the life cycle of matter and energy driving the formation of stars and the evolution of galaxies. BLAST Observatory will perform far-infrared polarimetry at sub-arcminute resolution, over large regions of sky, at wavelengths near the peak of the dust thermal spectrum that are inaccessible from the ground. Flying on a Super Pressure Balloon (SPB) from Wanaka, NZ, will allow for a 30-day flight with access to mid-latitude science targets and sufficient time for both legacy and shared-risk science programs.  We describe the scientific motivation for the BLAST Observatory and the proposed implementation, detailing the optical design of the instrument and the forecasting methods to predict its sensitivity
%both the technology necessary for balloon operation as well as the advances in submillimeter instrumentation enabling this 

%%\nbg{Note Giles' comment to rework text to be less like a proposal and BLAST specific and more generically promoting progress and tools for use by the community with BLAST as the specific case study.}

\end{abstract}



%\section{Introduction} \label{sec:intro}

\input{section_introduction}

\input{section_science}

\input{section_instrument}

\input{section_sensitivity}

\input{section_conclusions}

\input{acknowledgments}

\bibliography{main}{}
\bibliographystyle{aasjournal}

\appendix

\input{appendix_sensitivity_assumptions}



\end{document}

%% file: authors.tex
\collaboration{23}{(The BLAST Observatory Collaboration)}

\author[0000-0002-6362-6524]{Gabriele Coppi}
\affiliation{Department of Physics, University of Milano - Bicocca, Piazza della Scienza, 3 - 20126 Milano, Italy} 
\author[0000-0002-1940-4289]{Simon Dicker}
\affiliation{University of Pennsylvania, 209 S. 33rd St., Philadelphia, PA 19014, USA}

%%%%%%%%%%%%%%%%%%%%%%%%%%%%%%%%%%%
\author[0000-0002-4810-666X]{James E. Aguirre}
\affiliation{University of Pennsylvania, 209 S. 33rd St., Philadelphia, PA 19014, USA}
\author[0000-0002-6338-0069]{Jason E. Austermann}
\affiliation{National Institute of Standards and Technology, 325 Broadway, Boulder, CO 80305, USA}
\author[0000-0003-1263-6738]{James A. Beall}
\affiliation{National Institute of Standards and Technology, 325 Broadway, Boulder, CO 80305, USA}
\author[0000-0002-7633-3376]{Susan E. Clark}
\affiliation{Department of Physics, Stanford University, Stanford, CA 94305, USA}
\affiliation{Kavli Institute for Particle Astrophysics \& Cosmology, P.O. Box 2450, Stanford University, Stanford, CA 94305, USA}
\author[0000-0002-5216-8062]{Erin G. Cox}
\affiliation{Center for Interdisciplinary Exploration and Research in Astronomy (CIERA), Northwestern University, 1800 Sherman Avenue,  
Evanston, IL 60208, USA}
\author[0000-0002-3169-9761]{Mark J. Devlin}
\affiliation{University of Pennsylvania, 209 S. 33rd St., Philadelphia, PA 19014, USA}
\author[0000-0002-4666-609X]{Laura M. Fissel}
\affiliation{Queen's University, 64 Bader Lane, Kingson, ON, K7L 3N6, Canada}
\author[0000-0001-7225-6679]{Nicholas Galitzki}
\affiliation{University of Texas at Austin, 2515 Speedway, Austin, TX 78712, USA}
\affiliation{Weinberg Institute for Theoretical Physics, Texas Center for Cosmology and Astroparticle Physics, Austin, TX 78712, USA}
\author[0000-0001-7449-4638]{Brandon S. Hensley}
\address{Jet Propulsion Laboratory, California Institute of Technology, 4800 Oak Grove Drive, Pasadena, CA 91109, USA}
\author[0000-0002-2781-9302]{Johannes Hubmayr}
\affiliation{National Institute of Standards and Technology, 325 Broadway, Boulder, CO 80305, USA}
\author[0000-0002-9826-7525]{Sergio Molinari}
\affiliation{INAF-Istituto di Astrofisica e Planetologia Spaziali (IAPS), Via Fosso del Cavaliere 100, 00133 Roma, Italy}
\author[0000-0002-8307-5088]{Federico Nati}
\affiliation{Department of Physics, University of Milano - Bicocca, Piazza della Scienza, 3 - 20126 Milano, Italy}
\author[0000-0003-1288-2656]{Giles Novak}
\affiliation{CIERA and Department of Physics \& Astronomy, Northwestern University, Evanston, IL 60201, USA}
\author[0000-0003-1560-3958]{Eugenio Schisano}
\affiliation{INAF-Istituto di Astrofisica e Planetologia Spaziali (IAPS), Via Fosso del Cavaliere 100, 00133 Roma, Italy}
\author[0000-0002-0294-4465]{Juan D. Soler}
\affiliation{INAF-Istituto di Astrofisica e Planetologia Spaziali (IAPS), Via Fosso del Cavaliere 100, 00133 Roma, Italy}
\author[0000-0002-1851-3918]{Carole E. Tucker}
\affiliation{Cardiff University, School of Physics and Astronomy, Trevithick Buildings, Cardiff CF24 3AA, UK}
\author[0000-0003-2486-4025]{Joel N. Ullom}
\affiliation{National Institute of Standards and Technology, 325 Broadway, Boulder, CO 80305, USA}
\author[0000-0003-1246-4550]{Anna Vaskuri}
\affiliation{University of Colorado, Boulder, CO 80305, USA}
\author[0000-0003-2467-7801]{Michael R. Vissers}
\affiliation{National Institute of Standards and Technology, 325 Broadway, Boulder, CO 80305, USA}
\author[0000-0003-1678-5570]{Jordan D. Wheeler}
\affiliation{National Institute of Standards and Technology, 325 Broadway, Boulder, CO 80305, USA}
\author[0000-0002-4495-571X]{Mario Zannoni}
\affiliation{Department of Physics, University of Milano - Bicocca, Piazza della Scienza, 3 - 20126 Milano, Italy}

%% file: section_introduction.tex
\section{Introduction} \label{sec:intro}

Stratospheric balloon-borne platforms benefit from operating above more than 99.5\% of the Earth's atmosphere, allowing telescopes to observe regions of the electromagnetic spectrum that are not accessible to ground-based observatories. In particular, stratospheric balloon platforms are well suited to far-infrared and sub-mm astronomy \citep{pascale08,fissel16,lowe2020tng}. A recent advance in stratospheric ballooning is the advent of the Super Pressure Balloons (SPB).  Unlike standard zero-pressure balloons, which allow helium gas to escape when the pressure inside the balloon is higher than the surrounding atmospheric pressure, super pressure balloons are closed and maintain a nearly constant volume (and therefore buoyancy) even as the gas temperature changes. This technology enables long-duration flights from mid-latitude locations \citep[e.g.,][]{kierans2017,sirks2023}. 

The SPB platform will enable far-infrared/submillimeter observations at sub-arcminute resolution of most molecular clouds, covering many hundreds of square degrees of sky area.  By observing in the far-infrared and submillimeter these telescopes can observe the thermal emission from cold dust grains near the peak of the dust emission spectrum. Furthermore, by measuring the linear polarization of the emission from dust grains aligned with respect to the local magnetic field it is possible to map magnetic fields in the interstellar medium. These advances will  result in high sensitivity maps that address key outstanding questions about the interstellar medium, star formation, and the nature of interstellar dust.  In particular, these polarization maps can be used to (1) constrain the composition of interstellar dust, the seed material for the formation of stars and planets, (2) characterize interstellar turbulence, a key driver of energy transfer in galaxies, and (3) measure the relative importance of magnetic fields, turbulence, and gravity in the collapse of material to form stars. 

Sensitivity forecasts are crucial for proposing and executing new missions.  In this paper, we introduce a proposed SPB polarimetry mission, the Balloon-borne Large Aperture Submillimeter Telescope (BLAST) Observatory. We first describe  design characteristics the BLAST Observatory, then we provide a model for forecasting its sensitivity, and finally we compare the results with sensitivity specifications for other facilities, including recent, current, and planned instruments.  

In Section 2 of the paper we discuss the scientific motivation for the BLAST Observatory and how this drives the design.  Section \ref{sec:instrument} provides an overview of the technical specifications for BLAST Observatory, including the telescope, optics, detectors, and detector readout electronics.  Our detailed sensitivity model is presented in Section \ref{sec:sensitivity_model}, together with a complementary ``approximate model'' that is much simpler and serves to provide a ``sanity check'' on the results of the detailed model.  Section \ref{sec:sensitivity_model} also presents the comparisons with other facilities. We present our conclusions in Section \ref{sec:conclusions}.

%% file: section_science.tex
\section{Scientific Motivation}
\label{sec:science}

The following principal scientific objectives of the BLAST Observatory demonstrate the capability and potential scientific impact of a balloon-borne survey of far-infrared (far-IR) polarized dust emission.

\subsection{What are the physical properties of interstellar dust?}
\label{ssec:dust}

Models predict a systematic difference in the dust spectrum between total intensity and polarization depending on the composition(s) of dust present \citep[e.g.,][]{draine_fraisse_2009,Guillet2018,Hensley+Draine_2023,Ysard_2024}. The BLAST Observatory will measure the intrinsic properties of the dust polarization spectrum near its peak, where differences in dust models become pronounced \citep{Hensley2019}. 
 With its high sensitivity, the BLAST Observatory will produce wide-field polarization maps of diffuse dust in HI-clouds, translucent clouds with little molecular gas, and dozens of molecular clouds.
 Using these maps it will be possible to test whether one or more distinct varieties of dust exist in different environments and characterize the evolution of dust properties from diffuse gas to cold cores, revealing the mechanical, chemical, and thermodynamic processes that shape interstellar grains.
%\vspace{-2mm}

\subsection{What are the properties of magnetohydrodynamic (MHD) turbulence in the interstellar medium?}
\label{ssec:ism}
The interstellar medium (ISM) is a highly turbulent environment. Energy injected into the ISM, e.g. via supernovae or stellar feedback, cascades down a hierarchy of physical scales until it is dissipated.  
MHD turbulence structures the distribution of energy, magnetic fields, and material throughout the ISM, but its basic properties are poorly understood due to a dearth of data. The BLAST Observatory’s high-resolution maps of the diffuse ISM will allow us to measure the MHD turbulence power spectrum and test theories of how energy is dissipated in the ISM.

\subsection{What role do magnetic fields play in regulating star formation?}
\label{ssec:stars}
 Star formation in molecular clouds is controlled by the interplay between turbulence, gravity, and magnetic fields \citep[e.g.,][]{padoan14,krumholzANDfederrath2019,girichidis2020}. 
The BLAST Observatory will make the first statistical sample of high-resolution magnetic field maps covering entire molecular cloud complexes  and a large number of Galactic filamentary clouds. 
Observations will allow us to estimate magnetic field strength and energy density in clouds, filaments, and dense star-forming cores, and quantify the relationship between  magnetic field strength and the efficiency of star formation.

\subsection{Science-Driven Experimental Parameters}
\label{ssec:params}

The scientific goals of the experiment provide the top level requirements for the design. To resolve magnetic field morphology within filaments and cores in star-forming regions, as required for our star-formation science goals discussed in Section \ref{ssec:stars}, we require a resolution limit of less than 0.1\,pc for nearby ($d$\,$<$\,400\,pc) gas clouds, which implies that sub-arcminute resolution is needed. To test dust grain models (Section \ref{ssec:dust}) and map turbulent magnetic morphology in diffuse ISM clouds (Section \ref{ssec:ism}) we will require high signal-to-noise polarization maps of very faint dust columns ($\sigma_P\,\approx\,$0.1 MJy/Sr at 350\,$\mu m$).  Finally, to map entire nearby cloud complexes and large regions of the Galactic Plane the BLAST Observatory must be able to efficiently create large ($\sim$ 100 deg$^2$) polarization maps at multiple far-IR and sub-mm bands simultaneously.

%% file: section_instrument.tex
\section{Instrument Design} 
\label{sec:instrument}

\begin{figure}[t] 
%\vspace{-0.25in}
\begin{minipage}[c]{0.36\textwidth}
    \centering
    \hspace{-0.15in}
    \includegraphics[width=0.95\textwidth]{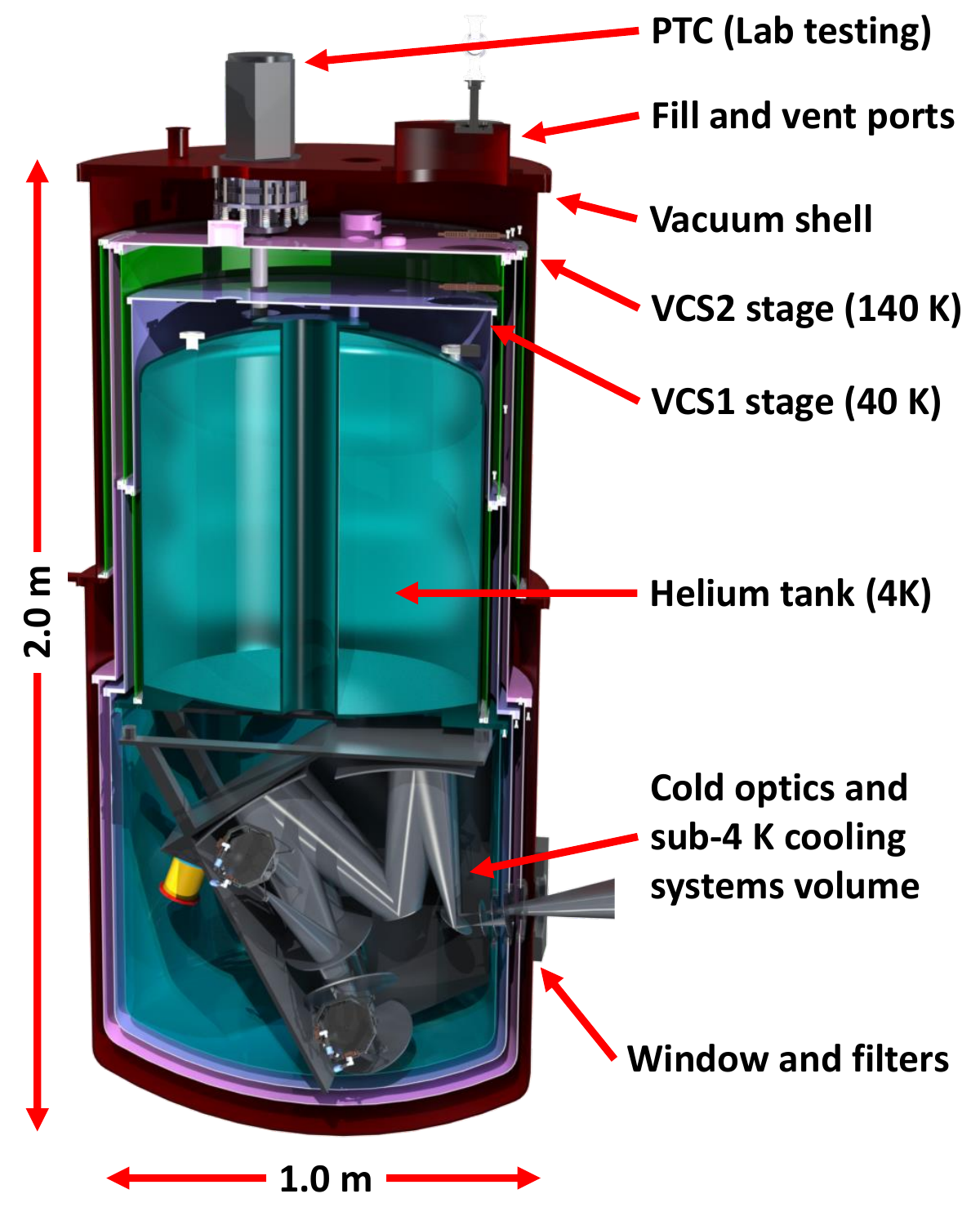}
\end{minipage}
\begin{minipage}[c]{0.5\textwidth}
    \centering
    \includegraphics[clip, trim=2cm 1cm 3cm 2cm, width=\textwidth]{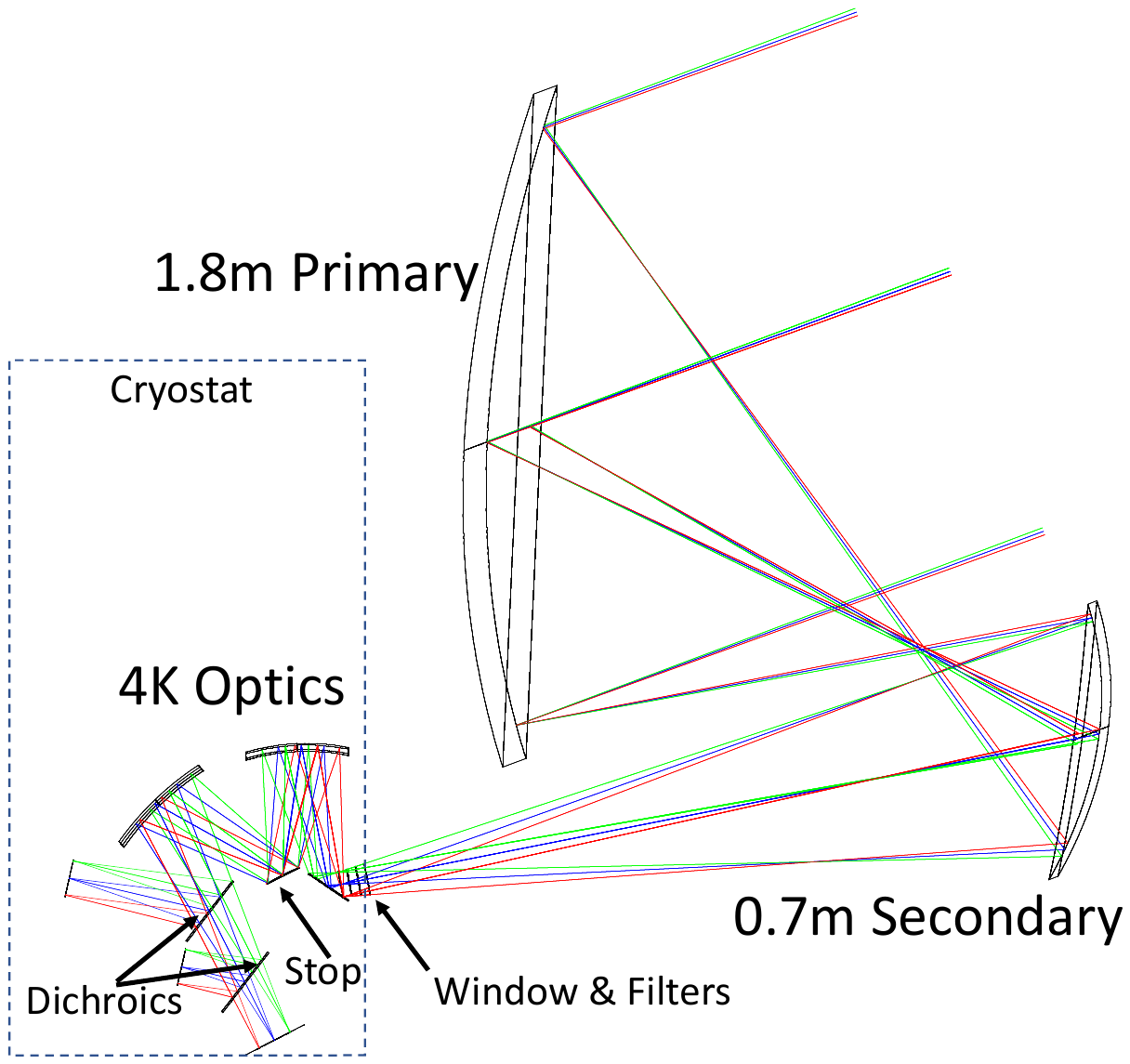} %{Figures/fig_optics_no_hwp.pdf}
\end{minipage}
    \caption{\small  Left: 
    A cross-section view of the BLAST Observatory camera
    which incorporates the cold optics and sub-4K stages built off the bottom of the $300$\,liter LHe tank. The LHe boil-off cools two vapor-cooled shields (VCS), which enable the projected 36 day hold time. Note the provision for pulse tube cooler (PTC) cooling during lab testing to reduce costs and risks associated with the LHe supply chain. The PTC will be removed for the deployment and flight phases. 
    Right: A simplified ray trace through the optics showing the different focal planes rotated to be in the plane of the page. Dichroic beam splitters pick off individual frequency bands so that the arrays observe the same 0.9$^\circ$\ diameter region on the sky. All focal planes are diffraction limited at all wavelengths. The dashed line shows the approximate location of the cryostat.}
    \label{fig:cryostat}
 \label{fig:optics}
\end{figure}

The BLAST Observatory instrument is described here in the context of improvements that maximize returns from using a balloon-based platform and feed into the subsequent sensitivity calculations. The BLAST Observatory will take full advantage of legacy technology and experience while concurrently utilizing advances in technology and capability to improve the design.  Our goal is to maximize the mapping speed of the instrument while making optimal use of the mid-latitude observations enabled by the Super Pressure Balloon (SPB). A summary of the instrument parameters is provided in Table \ref{tab:tel}. The principal changes to the instrument from previous generations are summarized as follows:

\begin{itemize}
    \item An off-axis  telescope design reduces the blockage and emission relative to an on-axis system, improving the projected mapping speed by 20\% (Figure \ref{fig:optics});
    \item A lower cryogenic base temperature of 100~mK and optimized TiN MKID detectors provide superior 1/$f$ noise performance compared to BLAST-TNG (Figure \ref{fig:dets1});
    \item Simultaneous observation with the same field-of-view (FoV) in three frequency bands, at 175~$\mu$m, 250~$\mu$m, and 350~$\mu$m optimizes our ability to discriminate polarized dust models and provides high resolution with our diffraction-limited optics (Figure \ref{fig:feeds1});
    \item Frequency multiplexed readout with lower power consumption than previous generations enables an increase in number of detectors to 8,274 -- 3 times the number in BLAST-TNG and 30 times the number in \blastp; 
    \item  A mid-latitude flight from New Zealand provides nighttime observing and more than 5 times the visible sky area compared to an Antarctic Long-Duration Balloon (LDB) flight.
\end{itemize}

\subsection{BLAST Observatory in context: Results from previous experiments}

The two previous generations of the experiment are worth comparing to BLAST Observatory to illustrate its advances in capability and to note the technical and scientific legacy that we are building on. These are \blastp, the \blast\ receiver \citep[e.g.,][]{devlin09} that was modified to include a polarimeter and flown in 2010 and 2012 from Antarctica, and BLAST-TNG, the polarimeter that flew from Antarctica in January 2020. \blastp\ provided the first high-resolution polarization maps covering entire star-forming clouds %\ref{fig:blastpol_velac}
\citep{fissel16,fissel2019,soler2017} and the first submillimeter polarization spectrum for a translucent cloud \citep{ashton2018}. Other \blastp\ results include the first whole-cloud polarization spectra for star-forming clouds \citep{gandilo16,santos17,shariff2019}. These results were highlighted in the National Academies of Sciences, Engineering, and Medicine's Astro2020 Decadal Report \citep[][their Fig.\ 6.2]{Decadal}.

BLAST-TNG featured large format MKID arrays in a long hold-time receiver, yielding dramatically improved mapping speed and enabling the first MKID observations from a NASA suborbital platform \citep{galitzki2016,lourie2018,lourie2018tel}. BLAST-TNG was launched in January 2020. Seconds after launch, the balloon collar impacted the payload with enough force to damage a carbon-fiber structural support. BLAST-TNG was able to operate for 15 hours, demonstrating performance for its detectors, gondola pointing systems, sub-Kelvin cryostat, and other subsystems \citep{coppi2020,lowe2020tng}. Unfortunately, the structural component ultimately failed and forced a flight termination.  The BLAST Observatory builds on this technical milestone, while adding improvements that provide a fourfold increase in mapping speed and a band at higher frequencies \citep[][Section \ref{sec:instrument}]{lowe2020blasto}. Additionally, the mid-latitude flight offers access to more than five times the sky area visible from Antarctica, enabling the expanded science goals of the BLAST Observatory. 

\begin{table}[]
\scriptsize
    \centering
    \begin{tabular}{lllcc}
        \toprule
        \toprule
         \textbf{Gondola} & Dimensions & \multicolumn{3}{l}{5.6\,m x 4.4\,m x 7.0\,m, Aluminum construction }\\
         \textbf{Telescope} & Temperature & \multicolumn{3}{l}{\textbf{day}: 260\,K \textbf{night}: 240\,K} \\
         & Primary diameter & \multicolumn{3}{l}{1.8\,m, 1.6\,m illuminated} \\
         & Primary RMS & \multicolumn{3}{l}{5\,$\mu$m (8\,$\mu$m)} \\
         & F-number & f/3.5 \\
         & Optical design & \multicolumn{3}{l}{Off-axis Gregorian} \\
%         & & & & \\
         \textbf{Cryostat} & L$^4$He Reservoir (L) & 300 \\
          & Cryogenics stages & \multicolumn{3}{l}{6: 140K, 40K, 4K, 2K, 0.270K, 0.1K} \\
          & Cryostat hold time (days) & 31 (36 Projected) \\
          & Sub-K technology & \multicolumn{3}{l}{Sorption Cooler + ADR} \\
%          & & & & \\
         \textbf{Detectors} & MKID quantum efficiency & 0.8 (0.7) \\
         & Feed-horn efficiency & 0.9 (0.7)\\
         & F-number at the detectors & f/5 \\ %this is approx - but good enough (srd)
         & Throughput & $A\Omega=\lambda^2$ \\
         & Central wavelength ($\mu m$) & \textbf{175} & \textbf{250} & \textbf{350} \\
         & Number of MKIDs & 3296 & 3296 & 1682 \\
        & Detector yield (included in MEV and CBE) & $>$80\% &  $>$80\%  &$>$80\% \\
         & Pixel size & $2.6f\lambda$ & $1.9f\lambda$ & $2f\lambda$ \\ %numbers assume 135mm array
%         & & & & \\
         \textbf{System} & Beam FWHM (arcsec) & 28 & 39 & 55 \\
         & Lyot stop fractional transmission & 0.6 & 0.6 & 0.6 \\
         & FoV per array (deg) & 0.9 \\
         & Filter widths ($\Delta\lambda/\lambda$) & 0.3 \\
%         & Observing Efficiency & 90$\%$ \\
         & Nominal scanning speed & 0.1$^{\circ}/$s \\
         \bottomrule
    \end{tabular}
    \caption{\small BLAST Observatory - telescope and receiver parameters. Table values are the Current Best Estimate (CBE) of the performance with Minimum Expected Values (MEV) performance in parentheses corresponding to the maximum allocated margins for all components in manufacturing and integration.}
    \label{tab:tel}
\end{table}

%%%%%%%%%%%%%%%%%%%%%%%%%%%%%%%%%%%%%%%%%%%%%%%%%%%%%%%%%%%%% 
\subsection{Telescope and Optics}

\label{sec:optics}

BLAST-Observatory will have an off-axis Gregorian optical design which has a number of advantages over on-axis systems, especially for submillimeter and millimeter telescopes.  However, an off-axis Gregorian design can suffer from degraded image quality.  This is easy to correct with a small ($\sim2^\circ$) tilt and a shift in the location of the secondary mirror allowing the lower sidelobes and detector loading inherent to off-axis designs while keeping a large field of view and low cross-polarization of on-axis telescope designs 

\begin{figure}
    \includegraphics[trim=5cm 6cm 5cm 2.9cm,clip,width=0.5\linewidth]{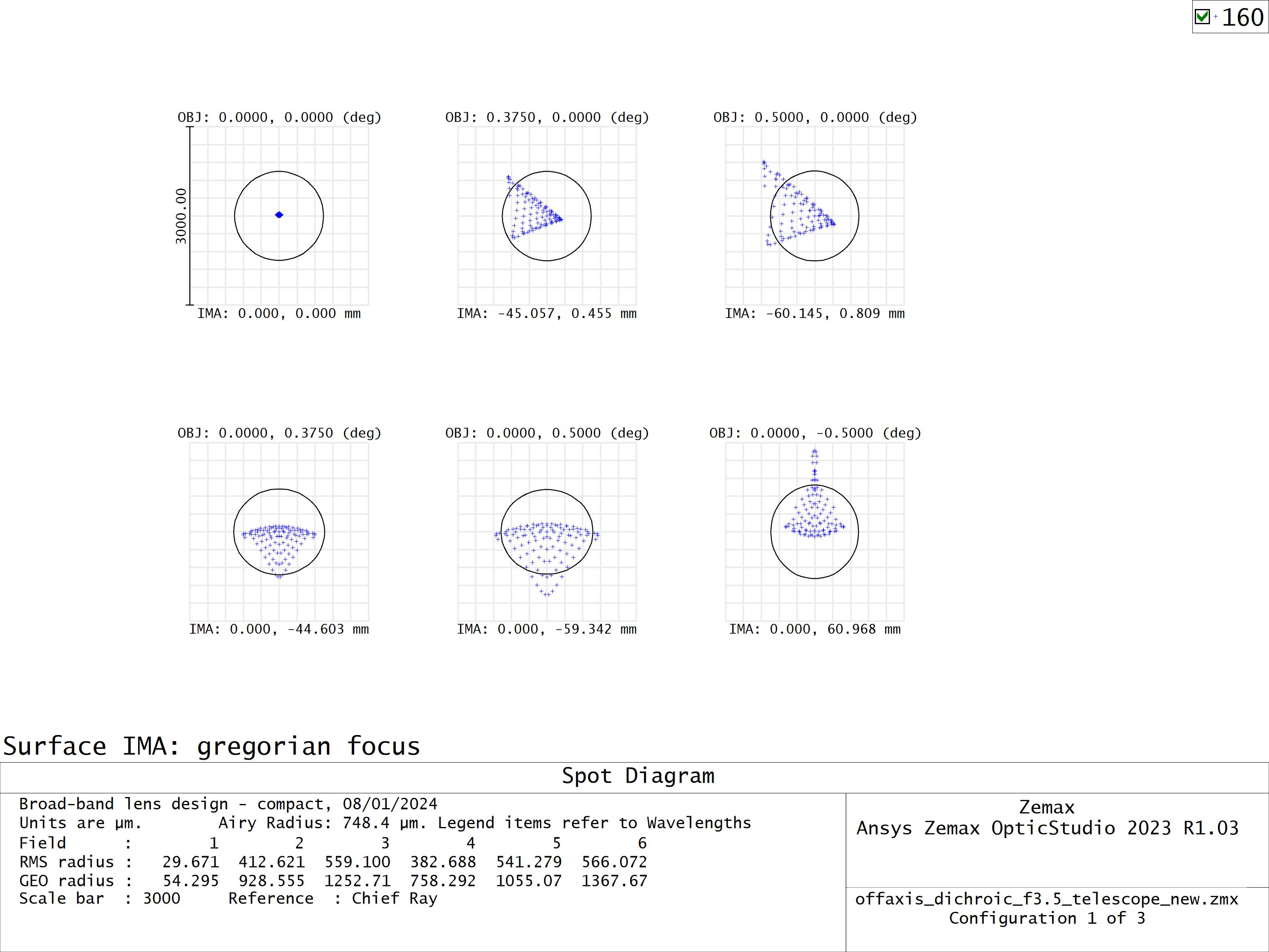}\hspace{1cm}
    \includegraphics[trim=5cm 6cm 5cm 2.9cm,clip,width=0.5\linewidth]{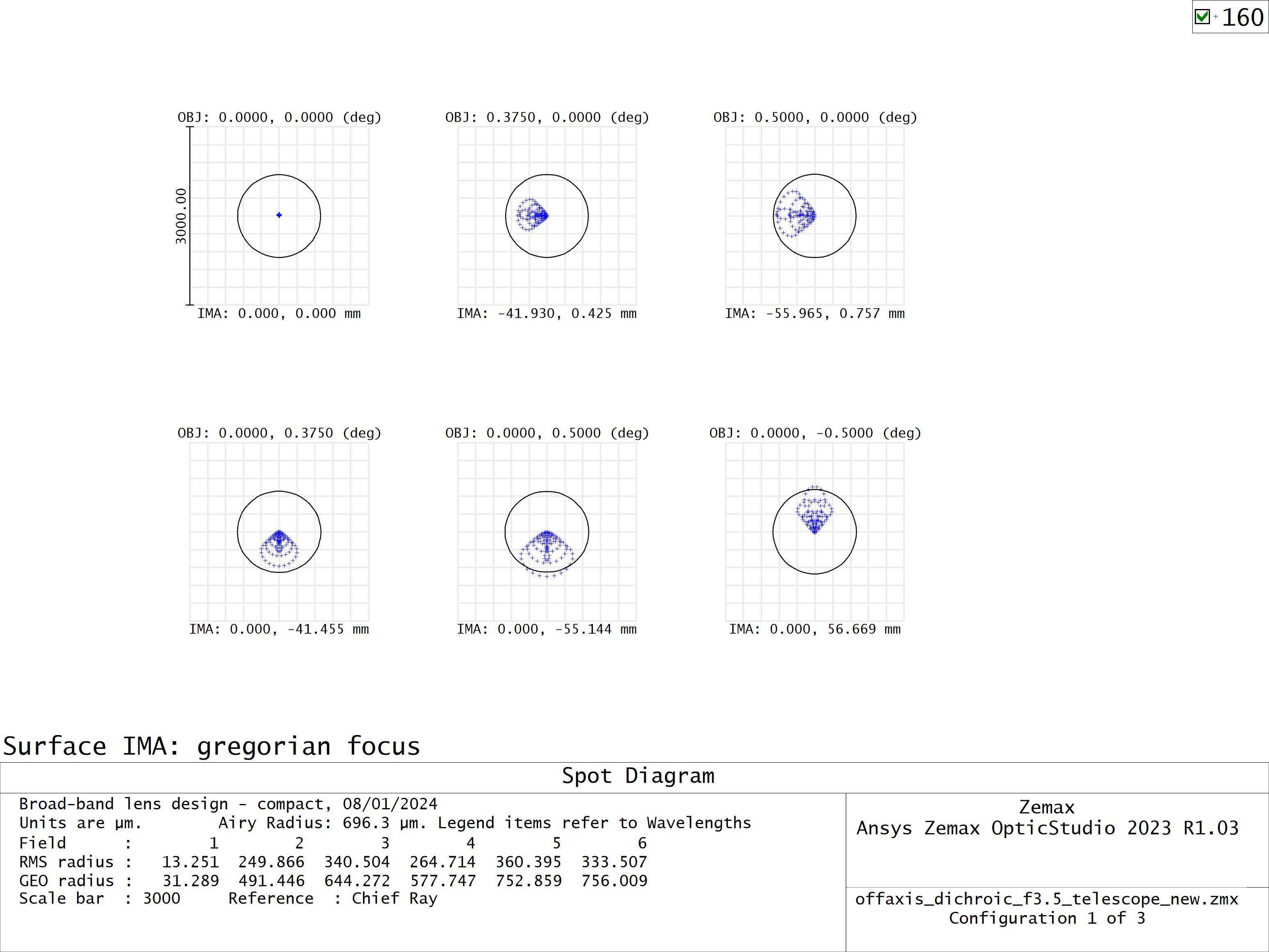}
    \caption{The BLAST Observatory telescope spot diagrams without (left) and with (right) a small tilt and decenter of the secondary mirror. The location of each spot diagram on the telescope's focal plane varies between the center and the edge, with the exact location given above each plot. These corrections produce image quality of an equivalent on-axis design. Individual rays are shown as blue points and the black circles represent the airy disk diameter. The relative locations and sizes of the mirrors are similar between the two designs allowing design studies to be reused. } \label{fig:new_telescope}
\end{figure}

The differences in mirror size and location with and without these corrections are small enough that the Monte Carlo design study results, described in Section \ref{sssec:tol}, are still applicable.  Spot diagrams with and without these corrections are shown in Figure~\ref{fig:new_telescope}. The higher image quality offered by this design results in simpler cold optics with a larger field of view and larger tolerances. The final design will have mirrors with large scale surface errors better than 5\,$\mu$m RMS, small-scale surface roughness surface roughness better than 0.5\,$\mu$m RMS, positional tolerances of 50\,$\mu$m, and focusing precision of 10\,$\mu$m.

%%%%%%%%%%%%%%%%%%%%%%%%%%%%%%%%%%%%%%%%%%%%%%%%%%%%%%%%%%%%%
\subsubsection{Cold optics}
\begin{figure}
\centering
    \includegraphics[trim=2cm 2.7cm 1cm 1cm,clip,width=0.47\linewidth]{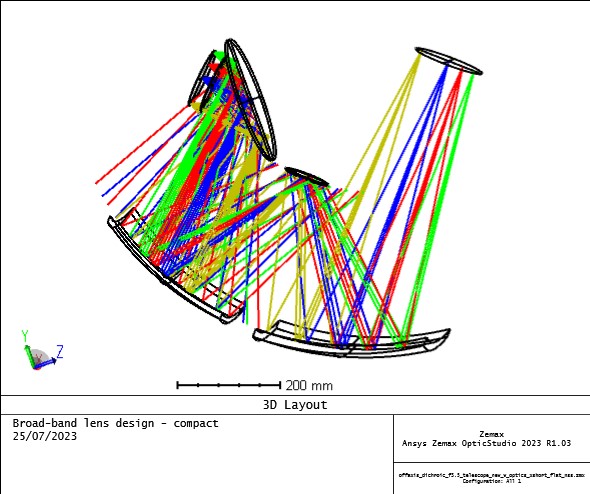} \hspace{0.5cm}
    \includegraphics[trim=6.5cm 9cm 6.5cm 1cm,clip,width=0.47\linewidth]{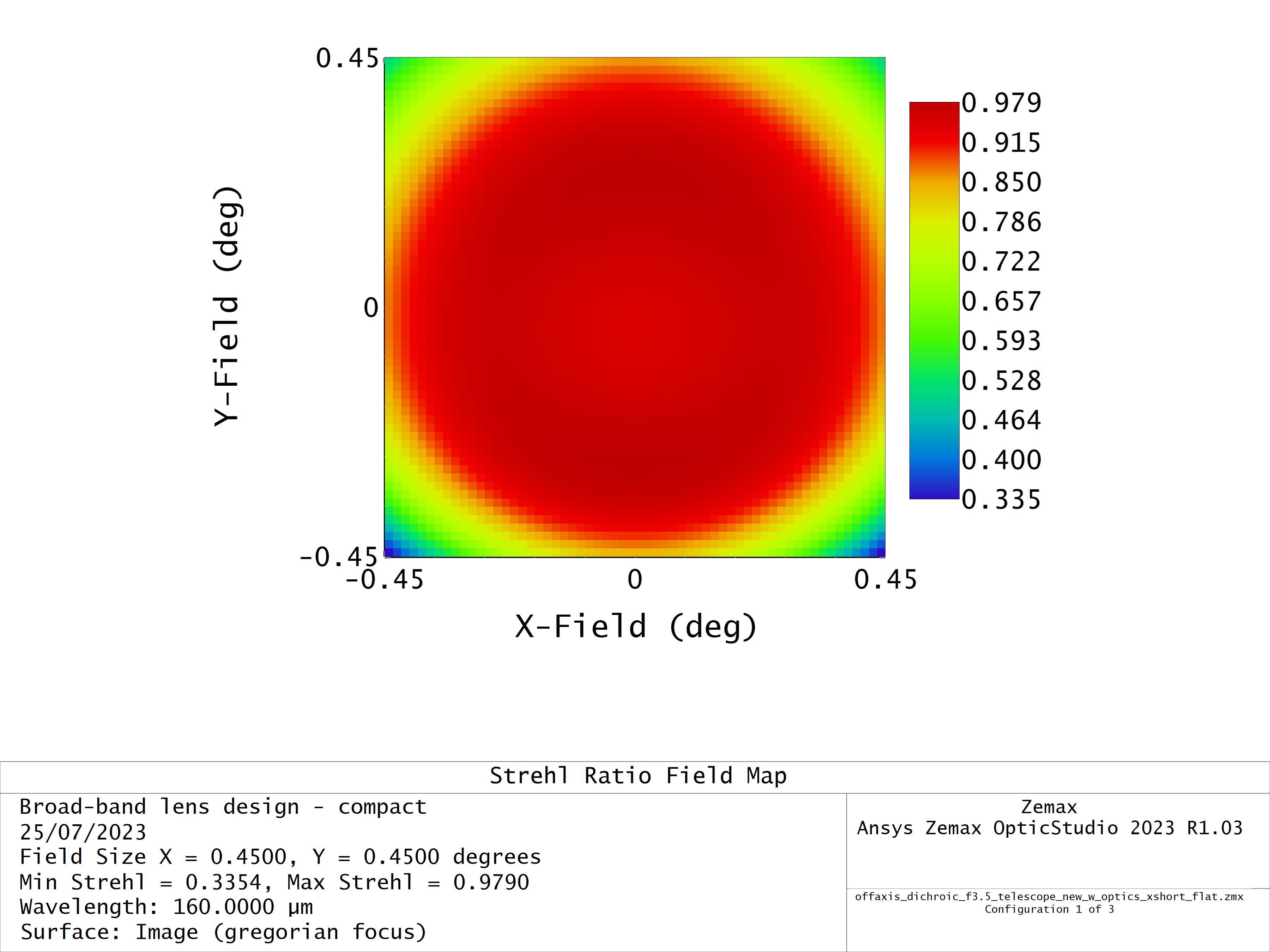}
    \caption{The cold optical design with rays outside the path from the primary included is shown at left.  Between the last mirror and the array is room for two dichroics.  There is ample room for absorbing baffles around the mirrors and the reflecting Lyot stop as, like BLAST, control of stray light within the optics cavity will be critical. The inside of the optical cavity will be lined with absorbing material.   It is simple to adjust the mirror and final focal length in order to fit the mechanical constraints of the cryostat and include extra room for cryogenic baffles.  This can be done without loss of image quality, as shown in the Strehl ratio map on the right at 160~$\mu$m. }\label{fig:coldOptics}
\end{figure}

The modified off-axis telescope design is coupled to a set of re-imaging optical surfaces cooled to 4\,K with a design based on an Offner relay (Figure \ref{fig:coldOptics}).  Constraints in the Optics studio design file allowed for adjustment of the final focal length and the spacing between the mirrors.  Because of the much larger throughput than previous BLAST telescopes, it was challenging to fit the components within the cryostat and not have mechanical interference.  As a result, while the mirrors can be enclosed on a single cold optical bench (surrounded by magnetic and optical shielding) a second optical bench at 45 degrees to the first will be needed to accommodate the arrays at images formed by dichroic beam splitters %\footnote{\gc{Dichroics are used to separate the different bands of Blast Observatory and observe the sky at the same time with all three arrays.}}
at locations that are easily accessible for cryogenic and electrical connections. Using dichroics that separate out the different frequency bands enables simultaneous observations of the same part of the sky in each band. This results in better overlap between maps made at each frequency and more efficient use of our focal plane. The focal planes are flat, telecentric, and diffraction-limited at all wavelengths. Although it is not planned, if it was found to be needed, a reflective polarization modulator could be placed at the position of the folding flat at the entrance to the cryostat. 

The cryostat design is easier to make taller than larger diameter which in-part motivates the configuration of the cold optic components. Additional space for optical components can be created by increasing the f-number which will in turn require a taller cryostat volume. Depending on the application or additional constraints, the design has a degree of flexibility to allow for re-optimization. 

\subsubsection{Optical Tolerances}
\label{sssec:tol}

A Monte Carlo analysis was carried out of the optical design with positions and tilts of all optical elements allowed to vary simultaneously.  The warm telescope was held to an accuracy of 50~$\mu$m with mirror tilts calculated as 50~$\mu$m side to center displacement. 
Without refocusing, image quality was maintained over 50 simulations where each element had a random position within the above tolerances.  In a second analysis we increased the variance in the telescope and receiver positions but allowed the secondary mirror to move to compensate. This showed that refocusing by actuating the secondary mirror could recover image quality at 160~$\mu$m to within 5\% of its design value. Results from a Monte Carlo tolerancing study are in Table~\ref{tab:tol}.  

All tolerances are well within a wide range of modern measurement techniques including laser trackers, measurement arms, and photogrammetry. The small angles correspond to tilts that move the edge of the mirrors by over 50\,$\mu$m (0.002$''$), easily within range of all of the above techniques.  The quoted accuracy of alignment of all elements from our Monte Carlo study will meet the requirements for the BLAST Observatory.  Although we do not expect large changes to occur after launch, should they happen, the actuated secondary will be able to recover image quality.  At the beginning of the flight, experience from BLAST has shown this can be done manually. However, we will also investigate automated procedures such as phase retrieval holography (sometimes known as OOF) which is currently used by other telescopes \citep{OOF}.

\begin{table}
\centering

\caption{\label{tab:tol} A summary of the tightest tolerances in the optical design with and without refocusing.  A Monte Carlo (MC) analysis where all parameters were allowed to vary simultaneously within these limits showed we can expect an increase in the wavefront RMS error of 2.9\% or 3.3\% with the looser limits when refocusing is allowed. The expected errors without refocusing include all thermal and gravitational distortions while those with refocusing reflect the accuracy to which we will be able to adjust the telescope. Note that many errors would change the absolute pointing of the telescope without degrading the image.  Our structure will be rigid enough that changes on short timescales are not expected and any long term changes such as thermal drifts can be taken out in map making.}
\begin{tabular}{llcccc}\hline
        &     &  MC Inputs (no refocus)   & Expected err & MC Inputs (With refocus) & Actuator accuracy  \\ \hline 
Primary & X / Y shift   & 100\,$\mu$m   & 50\,$\mu$m & 2000\,$\mu$m & \\ 
        & Tip / tilt    & 0.005$^\circ$ & 0.003$^\circ$  & 0.1$^\circ$ &  \\
        & Z             & 100\,$\mu$m   & 50\,$\mu$m        & 2000\,$\mu$m \\
        & RMS           & 10\,$\mu$m    & 5\,$\mu$m       & 10\,$\mu$m \\ 
Secondary & X / Y Shift & 100\,$\mu$m   & 50\,$\mu$m      & N/A   & 10\,$\mu$m \\ 
        & Tip / tilt    & 0.01$^\circ$  & 0.009$^\circ$ & N/A   &0.002$^\circ$ \\  
        & Z             & 100\,$\mu$m   & 50\,$\mu$m      & N/A   & 10\,$\mu$m \\
        & RMS           & 10\,$\mu$m    & 5\,$\mu$m       & 10\,$\mu$m  & \\
Cryostat & X / Y Shift  & 5\,mm         & 2.5\,mm         & 12\,mm & \\
        & Tip / tilt    & 1$^\circ$     & 0.5$^\circ$         &   1$^\circ$ & \\
        & Z             & 1\,mm         & 0.5\,mm         & 5\,mm & \\  
Cold optics & X / Y / Z shift   & 1\,mm & 0.9\,mm         & 2\,mm & \\ 
        & tilts         & 0.25$^\circ$  & 0.1$^\circ$         & 0.5$^\circ$ &\\
        &               &  \multicolumn{2}{l}{Wavefront from 0.031 to 0.052} &
                            \multicolumn{2}{l}{Wavefront from 0.028 to 0.062} \\\hline
\end{tabular}
\end{table}

\subsection{Detectors and Focal Plane} %{\normalsize (NIST)}} 
\label{sec:detectors}

\begin{figure}[t]
    \begin{center}
    \includegraphics[width=6.5in]{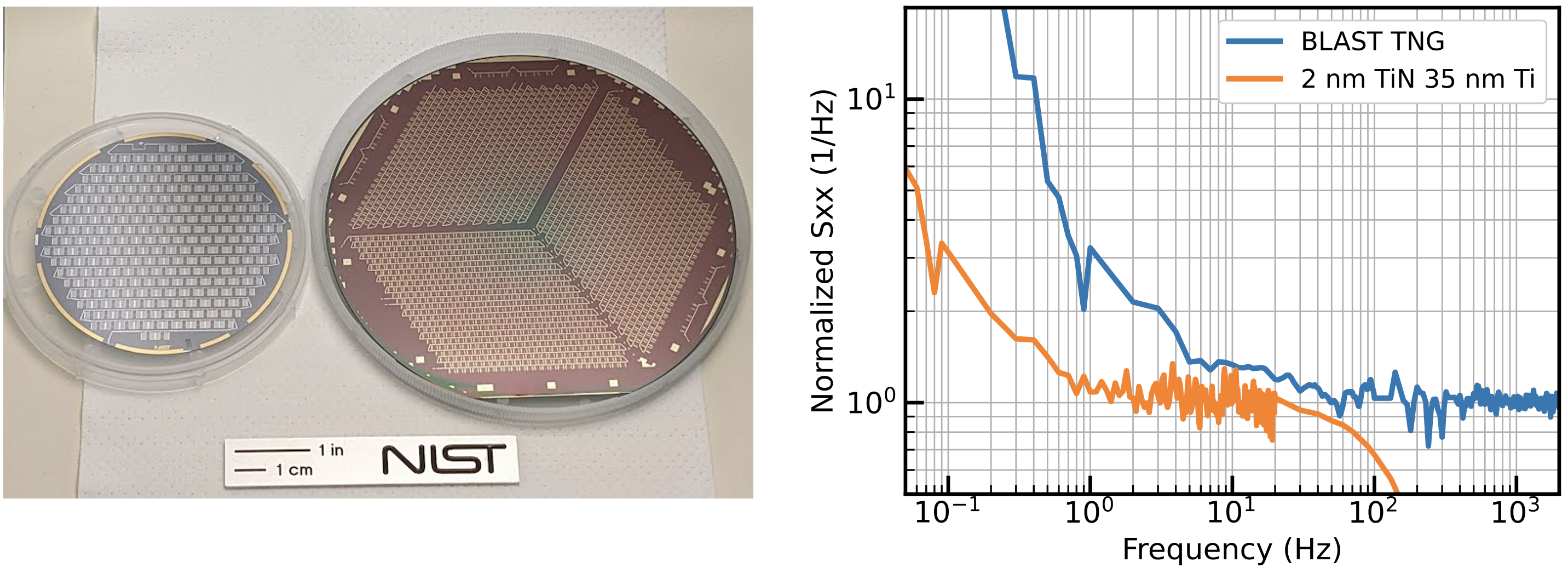}
    \end{center}
    \caption{\small 
    The detectors in BLAST Observatory will use the same general design and architecture as those demonstrated in-flight with BLAST-TNG. 
    Minor, low-risk changes to array fabrication will result in significant improvements in detector and instrumental performance.  
    {\textit{\textbf{Left:}}} Comparison of a 150~mm diameter array of detectors (right side of photo) compared to a spare array of 500~$\mu$m band BLAST-TNG detectors \citep{lourie2018} fabricated from a 100~mm diameter wafer. The 150~mm diameter demonstration array comprises 1728 pixels (3456 detectors) at a pixel pitch comparable to that intended for the highest density arrays of BLAST Observatory.
    {\textit{\textbf{Right:}}} Detector noise normalized to the photon-limited white noise level of an illuminated detector. A low-Tc TiN film that will nominally be used in BLAST Observatory (orange) exhibits significantly lower 1/$f$ noise and longer quasiparticle lifetimes than previous generation TiN-based KID devices.
    }
    \label{fig:dets1}
\end{figure}

BLAST Observatory will use the same fundamental approach to focal-plane design successfully flown on BLAST-TNG, 
while leveraging %NASA-supported 
technological advances at the National Institute of Standards and Technology to expand capability, increase sensitivity, and lower systematics.
This provides a vital testing platform for emerging technologies intended for future NASA missions while at the same time
being at a sufficient readiness level to ensure mission success. Specific improvements include: 
\begin{itemize}
    \item larger monolithic detector focal planes fabricated on 150~mm diameter wafers with a larger number of pixels and new processes for higher detector yield;
    \item recent advances in superconducting film materials for resonators allow the detectors to operate with significantly reduced low frequency (1/$f$) noise and enhanced sensitivity (Figure \ref{fig:dets1});
    \item replacement of direct-machined aluminum feedhorns with lithographically-defined silicon-platelet feedhorns capable of producing high-precision features, alignment, and uniformity that enable operation at higher frequencies while reducing systematic effects.
\end{itemize}

The BLAST Observatory focal planes will use polarization-sensitive microwave kinetic inductance detectors (MKIDs) \citep{day03} in a proven polarization-sensitive pixel design described in Refs. \cite{hubmay15} and \cite{dober16}.
This design has a demonstrated low cosmic ray cross-section and is fundamentally similar to detectors successfully flown in BLAST-TNG \citep{coppi2020}, deployed in TolTEC \citep{austermann18}, and produced for the first CCAT-prime continuum array \citep{choi22}.
Like TolTEC and CCAT-prime, the BLAST Observatory detectors will be fabricated on 150\,mm diameter wafers, allowing for large focal planes with more detectors and total system sensitivity than BLAST-TNG.  These arrays will also benefit from an increased detector yield using a post-fabrication resonator-editing technique  \citep{liu17}.

%% Detector Figure

The BLAST Observatory detectors will be further improved by incorporating newly-developed fabrication techniques. % and advances in superconducting materials.
While the high-$T_{\rm{c}}$ titanium nitride (TiN) detectors used in BLAST-TNG and TolTEC have successfully provided photon-noise limited performance at high signal frequencies \citep{hubmay15}, devices of this type have consistently exhibited excess 1/$f$ noise below a few Hz.  
Since then, Low-$T_{\rm{c}}$ TiN-based films have been demonstrated %(through NASA-supported development) 
to produce lower 1/$f$ knee frequencies \citep{wheeler19,wheeler22}.  
While many variations of Ti/TiN multilayers continue to be investigated, Fig.~\ref{fig:dets1} shows the normalized noise performance of an optically loaded MKID using a TiN bilayer consisting of 2~nm of TiN on top of 35~nm of Ti.  
This film exhibits a 1/$f$ knee $\sim$15$\times$ lower than the BLAST-TNG multilayer films, a $T_{\rm{c}}\sim 650$~mK, and significantly longer quasiparticle lifetimes that result in increased sensitivity of the MKID detector. 
Resonators based on this bilayer have maintained high quality factors, $Q\sim 200,000$ without optical loading, and a relatively high normal resistance of $\sim$7.5~ohm/$\square$ that plays a critical role in efficient coupling to waveguide.

Using this low-$T_{\rm{c}}$ film while operating at a lower base temperature allows the detectors to maintain photon-noise limited performance down to sub-Hz frequencies.
The BLAST Observatory is designed to provide a low bath temperature of $\sim$100~mK using an adiabatic demagnetization refrigerator (ADR) with a 300\,mK launch stage provided by a $^3$He sorption fridge. The cryostat uses a liquid helium architecture with vapor cooled shields and a pumped-pot to provide additional thermal stages at 2\,K, 4\,K, 40\,K, and 140\,K.
The detectors are protected from magnetic fields by a complete magnetic shield around the ADR magnet %(Sec.~\ref{sec:cryo}) 
and a high permeability shield around the 4\,K optics box. 
MKIDs routinely operate in this configuration in lab environments with ADRs,  with no effect on performance.  

Incident radiation is coupled to the detectors using silicon micromachined platelet feedhorn arrays \citep{britton10}.  
The silicon-based horns match the silicon substrate of the detector wafer and waveguide coupling components, allowing an all-silicon central focal plane unit.  
This configuration maintains precision alignment between room and cryogenic temperatures and helps eliminate microphonic noise by allowing for firm mounting of the detector array to the feedhorns with no need to allow for differential thermal contraction.
Lithographically-defined silicon platelets result in high precision horn features (typically better than $1\,\mu$m), which provide significant advantages in horn/waveguide symmetry and array uniformity compared to direct-machining that often suffers asymmetries and errors of several microns or larger.

\begin{figure}[t]
    \begin{center}
    \includegraphics[width=6.5in]{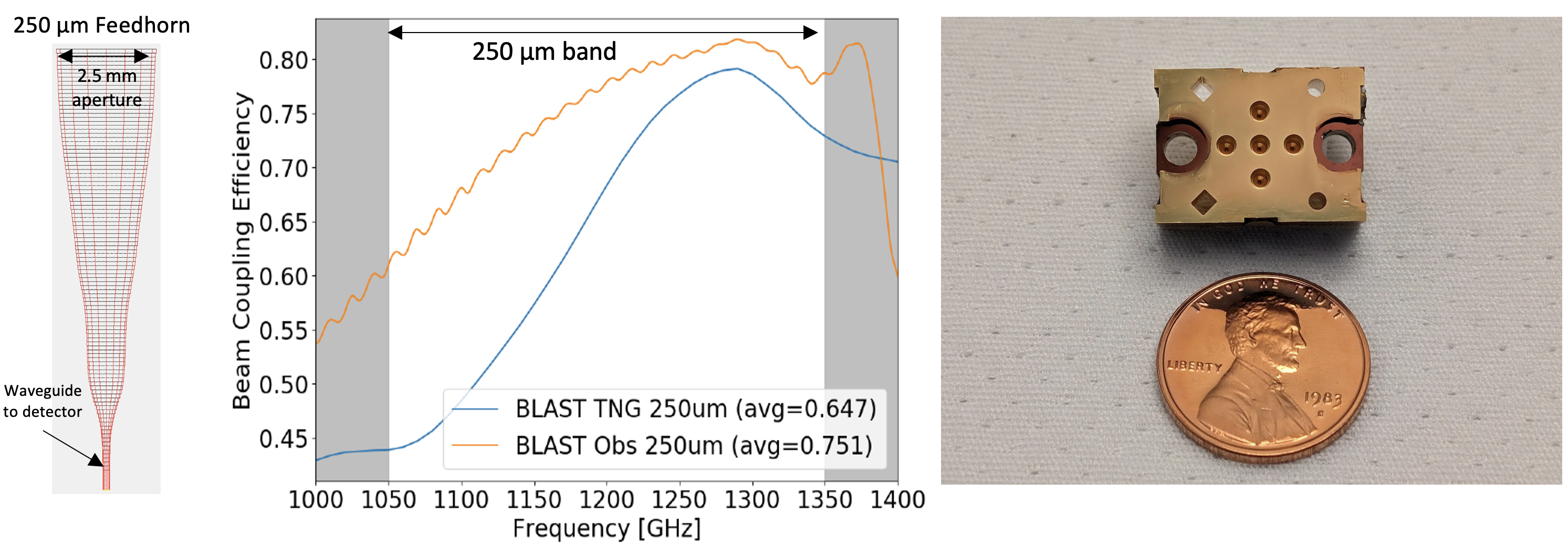}
    \end{center}
    \caption{\small 
    BLAST-TNG \citep{lowe2020tng} successfully demonstrated direct-machined aluminum feedhorns and waveguide in the far-infrared.  However, BLAST Observatory will baseline silicon-platelet feedhorns \citep{nibarger12} that provide many advantages including feedhorn precision, uniformity, performance, and are substrate-matched to the silicon-substrate detectors.  
    {\textbf{\textit{Left:}}}  Proposed platelet feedhorn design for the BLAST Observatory 250\,$\mu$m band.  
    {\textbf{\textit{Center:}}} Calculated beam coupling efficiency \citep{simon16} for a new platelet feedhorn design as compared to the previous BLAST-TNG aluminum horn, corresponding to in a 16\% increase in mapping speed.  
    {\textbf{\textit{Right:}}} Photograph of a prototype 5-pixel array of feedhorns optimized for the 350~$\mu$m band \citep{chapman22} demonstrating the scalability of silicon-platelet feedhorn fabrication to sub-mm wavelengths. 
    }
    \label{fig:feeds1}
\end{figure}

We have optimized the silicon feed profile through simulation for maximum performance in terms of beam efficiency, beam symmetry, reflection and cross-polarization. The resulting optimization for the 250\,$\mu$m band is shown in Figure~\ref{fig:feeds1} and represents a 16\% improvement in beam coupling compared to the metal BLAST-TNG feedhorn design of the same aperture size.  

Silicon platelet feedhorns are now a well-developed technology, having been used in a similar MKID experiments (TolTEC, CCAT) and multiple TES-based CMB projects (e.g., SPTPol, Advanced ACTpol, SPIDER). %\cite{austermann12,simon16,hubmayr16}).
The BLAST Observatory optics and pixel density allow for a feedhorn aperture and array density that is nearly identical to that already successfully demonstrated on the TolTEC 1100\,$\mu$m array \citep{wilson20}.  The most unique aspect of the BLAST Observatory feedhorn fabrication is the small waveguide diameter required, which has been successfully demonstrated \citep{lowe2020blasto} and a complete assembly has been prototyped in a relevant band (Fig.~\ref{fig:feeds1}).

\subsection{Detector Readout Electronics} %{\normalsize (ASU)}}
\label{sec:Readout}

BLAST-TNG demonstrated multiplexed readout of over 3000 detectors with five signal chains and warm electronics based on the FPGA-based Reconfigurable Open-Architecture Computing Hardware board (ROACH2). The BLAST Observatory will adapt the firmware developed for BLAST-TNG for the RFSoC ZCU111 board available from Xilinx. This board dramatically reduces the power to read out one detector over BLAST-TNG, from 59 to 14\;mW, enabling the readout of our 8274 detectors using only 112~W.

We only require four boards and eight readout lines (with 512 MHz of bandwidth each) and eight cryogenic low noise amplifiers (LNAs) for our three focal planes. Work is in progress to develop the firmware to demonstrate low power readout electronics for future space missions such as a far-IR probe-class mission and balloon missions like EXCLAIM \citep{Exclaim20} and TIM \citep{vieira20}. The firmware developed for these instruments will be leveraged for use in BLAST Observatory.

Environmental testing and power dissipation measurements of the RFSoC board have already been done.

The ability to read out more detectors per signal chain with lower power electronics  reduces both overall power consumption and the load on the cryogenic system from wires and amplifiers.
BLAST Observatory, as well as other balloon projects, will benefit from longer cryogenic hold times, reduced requirements on solar panels/battery capacity, and less waste heat to dissipate. While our focus is on using this technique for far-IR measurements, advances in FPGA-based readout will also directly benefit future infrared and X-ray observatories that require large numbers of superconducting detectors.

%% file: section_sensitivity.tex
\section{Sensitivity Model}\label{sec:sensitivity_model}

For BLAST Observatory, we developed two sensitivity models to estimate the performance of the instrument. For both models, we considered the case of photon-noise limited detectors. This assumption is based on the measurements of BLAST-TNG during its calibration phase in the 2020 Antarctic Campaign \citep{lowe2020tng}.
The first model that we developed is an analytical one that models each optical element as a frequency dependent component and considers the different temperature stage of those. The second model is included as a simpler sanity-check and uses several approximations including average values for frequency and temperature properties of the telescope. 

\subsection{Analytical Model: Theory}\label{subsec:analytical}

To introduce this model, we first estimated the optical power on a single pixel considering a single mode that excites the detector. To model the instrument, we considered a series of optical elements that can be considered as a gray-bodies. In this case, the optical power is then given by:

\begin{equation}
    \label{eq:popt}
    P_{opt}=\int_0^\infty W(\nu) \left[\sum_{i=0}^{N_{elem}}\left(\prod_{j=i+1}^{N_{elem}}\tau_j\right)\left(\sigma B_{s_i}(\nu, T_{s_i}) + \epsilon_i B_i( \nu, T_i)\right)\right] d\nu
\end{equation}

where $W(\nu)$ is the bandwidth of the focal plane, $\tau$ is the characteristic of each optical element of transmitting the radiation (reflectivity or transmissivity, depending on the component), $\epsilon$ is the emissivity of the element, $\sigma$ is the scattering coefficient of the element, $T_{s_i}$ is the temperature of the source that is scattered into the main beam and $T_i$ is the temperature of the component. In order to estimate the emissivity and optical efficiency of warm components, we use the assumptions described in Appendix \ref{sec:ass}. In particular, for the window we used the thickness-dependent model, given that the available data were taken for a thinner window. For the mirrors, we considered the scattering due to the roughness of the surface and described using a Ruze model. Finally, for components where the measurements were not available, we modeled the transmissivity as double logistic band-pass element with the plateau at a fixed value, extrapolated from the available spectra based on the filter material and thickness. Values for the optical elements are given in Table \ref{tab:optical_elements}.

$B(\nu,T)$ is the Planck spectral flux which, for a diffraction limited\footnote{In the diffraction limited case, $A\Omega = \lambda^2$ where $A$ is the collecting surface and $\Omega$ is the solid angle.} pixel, is given by:

\begin{equation}
    \label{eq:planckflux}
    B_{pixel}(\nu, T)=\frac{2h\nu}{e^{h\nu/\left(k_BT\right)}-1}
\end{equation}

Equation \ref{eq:popt} can also be extended to include the atmosphere as a \textit{virtual} optical element of the instrument. In particular, the properties of the atmosphere (transmissivity and temperature) are computed with the \textit{am - atmospheric model} software \citep{Paine2023} using the data from the MERRA-2 reanalysis\footnote{https://gmao.gsfc.nasa.gov/reanalysis/MERRA-2/}. Finally, the background power is given by the dust modeled as a modified black-body with a temperature of $19.7$ $K$:

\begin{equation}
    \label{eq:dust}
    BB_{dust} = A_{\nu_{ref}}\left(\frac{\nu}{\nu_{ref}}\right)^{\alpha}BB(\nu, T_d)
\end{equation}

where $A_{\nu_{ref}} = 0.002$ is the amplitude at the reference frequency $\nu_{ref} = 353$ GHz and $\alpha = 1.51$ is the spectral index \citep{Planck2018_XI}.

To estimate the noise on the detectors, we followed the formulation presented in \cite{Lamarre1986} where the noise equivalent power (NEP) on a single pixel is given by: 

\begin{equation}
    \label{eq:noise}
    NEP^{2}_{phot_{pixel}} = 2h\int_0^\infty W(\nu)\nu Q P(\nu) d\nu + 2 \int_0^\infty Q^2 P^2(\nu) d\nu \quad [W/\sqrt{Hz}]
\end{equation}

where $Q$ is the quantum efficiency of the detector, $P(\nu)$ is the optical power on the detector computed using Equation \ref{eq:popt}. This formulation includes both the shot and bunching noise as the first and second components in the equation, respectively. 
Both the $P_{opt}$ and the $NEP^2$ are computed per pixel. For an experiment like BLAST Observatory, where each pixel contains two linear polarization sensitive detectors, we need to divide by a factor of 2 to get the optical power and the noise on each detector. 

The $NEP$ is a useful figure of merit for understanding the performance of the instrument, but for a submillimiter experiment it is important to introduce other metrics to evaluate the noise performances. In particular, we introduce the \textbf{Noise Equivalent Flux Density} for a single detector observation as:

\begin{equation}
    \label{eq:NEFD_det}
    NEFD_{det}^{sky} = NEP_{det}\frac{10^{26}}{\sqrt{2}\pi r^2\eta\Delta\nu} [W \cdot \sqrt{s}].
\end{equation}

The factor of $\sqrt{2}$ arises when converting from $Hz$ to $s$, the $10^{26}$ is the conversion from W to Jy, $r$ is the radius of the primary mirror, $\Delta\nu$ is the frequency bandwidth, and $\eta$ is the total optical efficiency of the instrument including the quantum efficiency of the detectors. The NEP for a single detector is computed as $\sqrt{2}\cdot NEP_{pixel}$. %It is defined as \emph{$^{sky}_{det}$} since the NEFD is a result of single detector observation.
Given Equation \ref{eq:NEFD_det}, it is possible to compute the on-sky NEFD for the array as:

\begin{equation}
    \label{eq:NEFD_arr}
    NEFD_{arr}^{sky} = \frac{NEFD_{det}^{sky}}{\sqrt{N_{det} Y}}
\end{equation}

where $N_{det}$ is the number of detectors and $Y$ is the yield of the array.

\begin{table}[]
    \centering
    \addtolength{\leftskip}{-2cm}
    \begin{tabular}{lccccccc}
        \toprule
        Components & Temperature (K) & \multicolumn{3}{c}{Optical Efficiency} & \multicolumn{3}{c}{Emissivity}  \\
        & & 175$\mu$m & 250$\mu$m & 350$\mu$m & 175$\mu$m & 250$\mu$m & 350$\mu$m \\
        \midrule
        Primary & 270 & 0.77 & 0.78 & 0.85 & $1.07\cdot 10^{-2}$ & $1.07\cdot 10^{-2}$ & $1.07\cdot 10^{-2}$\\
        Secondary & 270 & 0.79 & 0.88 & 0.93 & $1.07\cdot 10^{-2}$ & $1.07\cdot 10^{-2}$ & $1.07\cdot 10^{-2}$\\
        Warm Filter & 270 & 0.88 & 0.88 & 0.88 & $9.06\cdot10^{-4}$ & $9.13\cdot10^{-4}$ & $9.11\cdot10^{-4}$\\
        Window & 270 & 0.96 & 0.95& 0.95 & $3.92\cdot10^{-2}$ & $1.77\cdot10^{-2}$ & $4.26\cdot10^{-3}$\\
        IR Filter & 270 & 0.88 & 0.93 & 0.96 & $2.07\cdot 10^{-3}$ & $1.35\cdot 10^{-3}$ & $9.82\cdot 10^{-4}$\\
        IR Filter & 155 & 0.82 & 0.89 & 0.93 & $1.88\cdot 10^{-3}$ & $1.26\cdot 10^{-3}$ & $9.47\cdot 10^{-4}$ \\
        IR Filter & 45 & 0.82 & 0.89 & 0.93 & $1.88\cdot 10^{-3}$ & $1.26\cdot 10^{-3}$ & $9.47\cdot 10^{-4}$ \\
        LP Filter & 45 & 0.91 & 0.91 & 0.92 &  $1.24\cdot 10^{-3}$ & $1.20\cdot 10^{-3}$ & $9.62\cdot 10^{-4}$ \\
        IR Filter & 4 & 0.81 & 0.89 & 0.93 & $1.82\cdot 10^{-3}$ & $1.25\cdot 10^{-3}$ & $9.45\cdot 10^{-4}$ \\
        LP Filter & 4 & 0.84 & 0.84 & 0.85 & $9.99\cdot 10^{-3}$ & $9.98\cdot 10^{-3}$ & $9.99\cdot 10^{-3}$ \\
        LP Filter & 4 & 0.89 & 0.89 & 0.89 & $2.20\cdot 10^{-3}$ & $1.33\cdot 10^{-3}$ & $9.22\cdot 10^{-4}$ \\
        M3 & 4 & 0.99 & 0.99 & 0.99 & $1.07\cdot 10^{-2}$ & $1.07\cdot 10^{-2}$ & $1.07\cdot 10^{-2}$\\
        M4 & 4 & 0.99 & 0.99 & 0.99 & $1.07\cdot 10^{-2}$ & $1.07\cdot 10^{-2}$ & $1.07\cdot 10^{-2}$\\
        M5 & 4 & 0.99 & 0.99 & 0.99 & $1.07\cdot 10^{-2}$ & $1.07\cdot 10^{-2}$ & $1.07\cdot 10^{-2}$\\
        Dichroic 1 & 4 & 0.84 & 0.84 & 0.84 & $6.00\cdot 10^{-2}$ & $6.00\cdot 10^{-2}$ & $6.00\cdot 10^{-2}$ \\
        Dichroic 2 & 4 & NA & 0.84 & 0.84 & NA & $6.00\cdot 10^{-2}$ & $6.00\cdot 10^{-2}$\\
        Bandpass Filter & 0.15 & 0.83 & 0.85 & 0.91 & $7.00\cdot 10^{-2}$ & $6.72\cdot 10^{-4}$ & $4.89\cdot 10^{-4}$\\
        \bottomrule
    \end{tabular}
    \caption{Optical Elements and their temperature and mean optical efficiency of the component in-band. \textbf{For the mean value, we defined the band as centered in the nominal value and the edges are set at half-power of the value of the nominal value.}}
    \label{tab:optical_elements}
\end{table}

The last Figure of Merit, that we want to use to estimate the performance of BLAST is the RMS of an observed map. This can be calculated as: 

\begin{equation}
    \label{eq:sigma_map}
    \sigma_{I} = \sqrt{\frac{A}{MS\cdot t}}
\end{equation}

where $A$ is the survey area in square degrees and $t$ is the observation time in seconds and $MS$ is the mapping speed of the instrument. The latter is defined as the number of square degrees of sky that can be mapped to a 1 Jy noise level (so with a variance of 1 Jy$^2$ ) in beam-sized map pixels, in a second. 
With this definition, each array has the following mapping speed:

\begin{equation}
    \label{eq:MS}
    MS = \frac{1}{N_{beams}NEFD_{sky_{arr}}^{2}}
\end{equation}

where $N_{beams}$ is the number of beams in a square degree, which can be computed as:

\begin{equation}
    \label{eq:Nbeams}
    N_{beams} = \frac{1}{FWHM^2*4/\pi/\ln{2}}.
\end{equation}

Until now, we have not considered the polarization sensitivity of the instrument. Given the nature of the BLAST Observatory, it is important to define also the figure of merits for a polarized signal. To properly estimate the noise in polarization, it is necessary to introduce the polarization efficiency, $\eta_{P}$, that encapsulates the instrumental polarization properties. Moreover, we need to consider that each of the detectors in BLAST Observatory is sensitive only to a single polarization, doubling the noise for a single detector. With these considerations, it is possible to formalize the NEP, NEFD, and RMS of the polarized map. The NEP is given by

\begin{equation}
    \label{eq:NEP_pol}
    NEP_{P} = \frac{\sqrt{2}NEP}{\eta_P}.
\end{equation}

Similarly, the NEFD from Equation \ref{eq:NEFD_arr} becomes:

\begin{equation}
    \label{eq:NEFD_pol}
    NEFD_{P} = \frac{\sqrt{2}NEFD}{\eta_P}.
\end{equation}
Finally, it is possible define the error on the polarized flux as:

\begin{equation}
    \label{eq:sigma_P}
    \sigma_{p} = \frac{\sqrt{2}\sigma_{I}}{\eta_P}.
\end{equation}

where the increase by $\sqrt{2}$ arises because the detectors must measure two quantities (Stokes Q and Stokes U) rather than just one (Stokes I).  (Note that from the point of view of error propagation, the differences between detector signals that are used to calculate Q and U are no different than the sums of signals used to calculate I.)

\begin{table}[]
\vspace{-0.2in}
    \scriptsize
    \centering
    \begin{tabular}{lcccr}
        \toprule
        \textbf{Parameter} & \multicolumn{3}{c}{\textbf{Value}} &  \textbf{Notes}\\
        \midrule
        $T$ & \multicolumn{3}{c}{270 K} & assigned to \textit{warm} components\\
        Central Wavelength & 175 & 250 & 350 & in microns \\
        Bandwidth & $30\%$ & $30\%$ & $30\%$ & $\Delta \lambda / \lambda$ \\
        $\eta_P$ & $0.9$ & $0.9$ & $0.9$ & from \cite{Galitzki2014} \\ 
        $t_{cold}$ & $0.38$ & $0.44$ & $0.53$ & from model filters\\
        $\epsilon_{warm}$ & 0.065 & 0.063 & 0.037 & from model atmosphere, mirrors, window\\
        $t_{warm}$ & $0.47$ & $0.63$ & $0.73$ & from model atmosphere, mirrors, window\\
        $(A\Omega/\lambda_0^2)$ & \multicolumn{3}{c}{1} & $\lambda_0 = c/\nu_0$ \\
        $R$ & \multicolumn{3}{c}{0.8 m} & from Table \ref{tab:tel}\\
        $FWHM$ arcsec & 28 & 39 & 55 & from Table \ref{tab:tel}\\
        $N_{det}$ & 3296 & 3296 & 1682 & from Table \ref{tab:tel}\\
        $Yield$ & \multicolumn{3}{c}{0.8} & from Table \ref{tab:tel}\\
        \cline{1-5}
    \end{tabular}
    \caption{\small Input parameters for the approximate model. The emissivities and trasmissivities are computed from the optical elements' frequency dependent curves. In particular, we first compute the frequency dependent curve for the entire group (so {\em cold} or {\em warm}) and then we compute the mean in the $\Delta \lambda / \lambda$ interval.}
    \label{tab:sanity_check}
\end{table}

\subsection{Approximate Model}
As a simple check on the these results from our sensitivity model, we next developed an approximate method to estimate the sensitivity figures of merit.  This is meant to provide a quick way for the reader to (approximately) verify our results and to understand how the mapping speeds depend on the various specifications we have assumed.  A key simplification comes from treating each optical element including the atmosphere as either being at 270 K (\textit{warm}) or 0 K (\textit{cold}), depending on whether its actual temperature in the full sensitivity model is above or below 150 K.  Also, we assume that the filter passband has a box function shape with cut-on and cut-off centered at the nominal wavelength and a width equal to $30\%$ for each band. We use the mid-band frequency to calculate the photon energy, and we employ the Rayleigh-Jeans approximation to the Planck function.  Finally, we treat the photon noise as \textit{shot noise}, ignoring corrections for Bose statistics, and we assume the detectors have perfect absorptivity.

In our simplifed treatment, the optical power incident on a single (polarized) detector can be written in terms of the emissivity $\epsilon_{warm}$ and transmission $t_{warm}$ of the 270 K elements, the transmission $t_{cold}$ of the 0 K elements, the product $A\Omega$ of the area $A$ of primary mirror and area $\Omega$ of telescope beam, and the mid-band frequency and passband values $\nu_0$ and $\Delta\nu$, as follows:

\begin{equation}
    \label{eq:power_approx_1}
     P = \frac{1}{2}\left(\frac{2kT\nu_0^2}{c^2}\right)\epsilon_{warm}(\Delta\nu)(A\Omega)t_{cold} t_{warm}. 
\end{equation}

The factor of one half is needed because we consider only one polarization component.  (The parameters $t_{cold}$ and $t_{warm}$ are referenced to a single polarization component.)  As noted above, we will set $T$ = 270 K.  If we define $\lambda_0$ = $c^2\nu_0^{-2}$, we can write this as

\begin{equation}
    \label{eq:power_approx_2}
     P =  kT\epsilon_{warm}(\Delta\nu)t_{cold} t_{warm}(A\Omega / \lambda_0^2)
\end{equation}

where we have grouped $A\Omega$ and $\lambda_0^2$ at the end because we will set $A\Omega$ = $\lambda_0^2$.   

Under the simple assumptions of our approximate calculation, the amplitude spectral density $NEP_{phot}$ of the photon shot noise is given by

\begin{equation}
    \label{eq:NEP_approx}
     NEP_{phot} = \left(2 h\nu_0 P \right)^{0.5}
\end{equation}

which is constant with frequency (white noise).  This quantity is usually expressed in units of $W Hz^{-0.5}$.  

\begin{figure}
    \centering
    \includegraphics[width=0.7\textwidth]{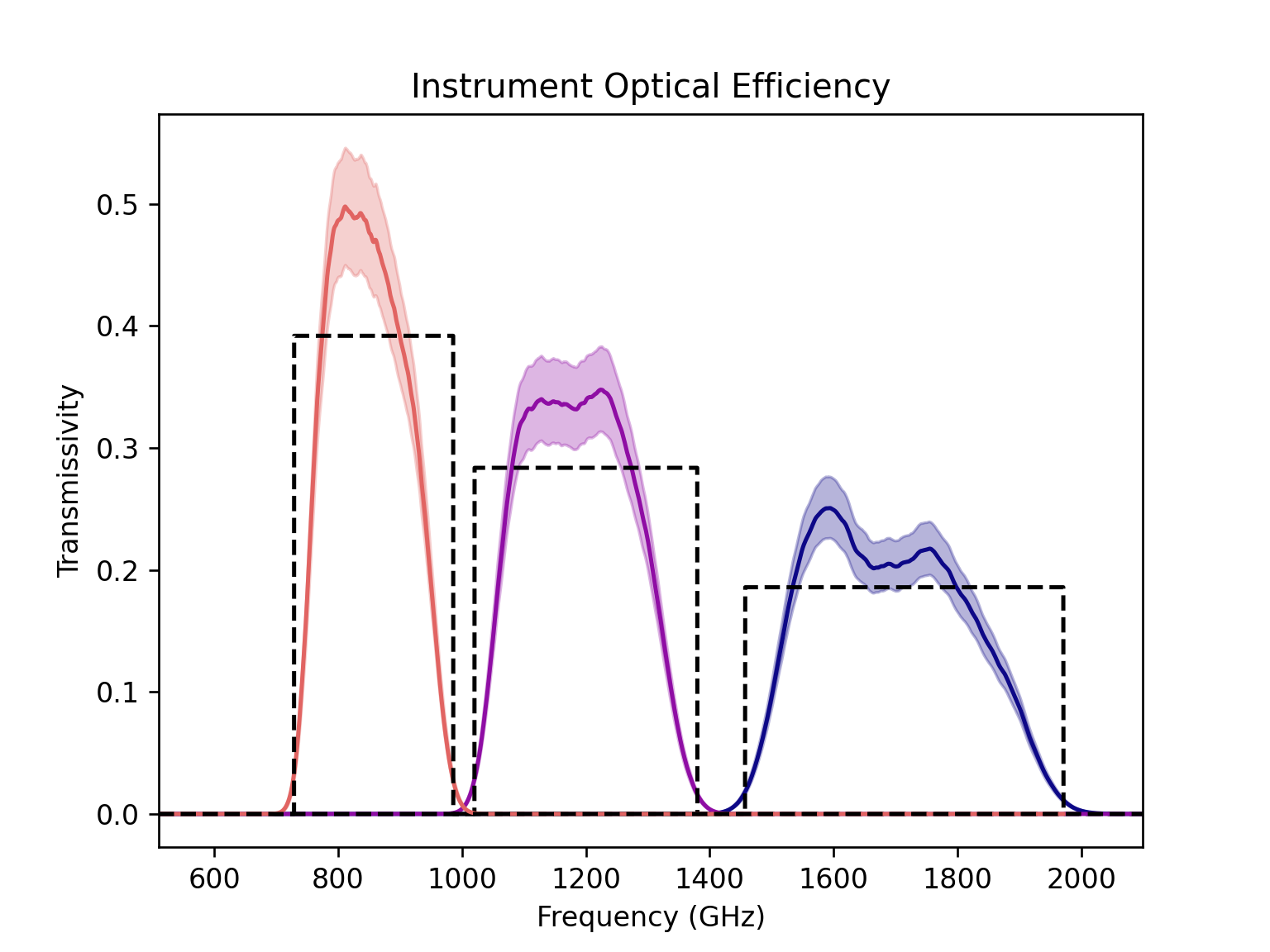}
    \caption{BLAST Observatory Optical Efficiency. The shaded regions represent the 10th to 90th quantile distribution from the Monte Carlo analysis for each band. The dashed lines represent the instrumental optical efficiency used for the approximate model.}
    \label{fig:opticalefficiency}
\end{figure}

Next consider the quantity $P_{SNR=1}$, defined as the power incident on a single detector that will yield an SNR of unity for an integration time of duration $t$. Because a 1 $s$ integration corresponds to a bandwidth of 0.5 $Hz$ \citep{Richards94}, we obtain the following expression for $P_{SNR=1}$:

\begin{equation}
    \label{eq:power_for_unity_SNR_approx}
     P_{SNR=1} = \frac{NEP_{phot}}{\sqrt{2}} t^{-0.5}
\end{equation}

which can be rewritten as

\begin{equation}
    \label{eq:power_for_unity_SNR_approx_rewritten}
     P_{SNR=1} = \left(h\nu_0\right) \left(\frac{P}{h\nu_0}\right)^{0.5} t^{-0.5}
\end{equation}

where we have rearranged terms to make it obvious that this equation can be derived directly by considering an integration time of 1 $s$ and applying root-N statistics.  From this it follows that the {\em total} flux density required to obtain (in time $t$) SNR=1 detections at {\em each} of the two orthogonally polarized detectors of a given feedhorn can be written as

\begin{equation}
    \label{eq:flux_for_unity_SNR_approx}
     FD_{SNR=1} = \frac{P_{SNR=1}}{\pi R^2 \Delta\nu \left( t_{cold} t_{warm} \right)} = \frac{NEP_{phot}t^{-0.5}}{\sqrt{2}\pi R^2 \Delta\nu \left( t_{cold} t_{warm} \right)} 
\end{equation}

where R is the effective radius of the primary mirror.  Finally, multiplying by $t^{0.5}$ we obtain the Noise Equivalent Flux Density $NEFD_{phot}$, which is defined as the total flux density required to obtain SNR=1 at each of two polarized detectors multiplied by the square root of the integration time assumed:

\begin{equation}
    \label{eq:NEFD_approx}
     NEFD_{phot} = \frac{NEP_{phot}}{\sqrt{2}\pi R^2 \Delta\nu \left( t_{cold} t_{warm} \right)} = \frac{\left(h\nu_0 kT\epsilon_{warm}(A\Omega / \lambda_0^2)\right)^{0.5}}{\pi R^2\left((\Delta\nu)t_{cold} t_{warm}\right)^{0.5}}.
\end{equation}

The units for this quantity are traditionally written as $\mathrm{mJy~ s}^{0.5}$.

Finally, the achievable uncertainty in total flux for mapping one sq. deg. in 1 hour, $\sigma_{I}$ is found from

\begin{equation}
    \label{eq:mapping_speed_total_intensity_approx_1}
     \sigma_{I} = \left(NEFD_{phot}\right) (1\,h)^{-0.5}  \left(\frac{1\,deg^2}{\Omega}\right)^{0.5} (N_{det}Yield)^{-0.5} \left(\frac{\Omega}{SR}\right)^{-1}
\end{equation}

where we have divided by the beam area $\Omega$ so that we may express the achievable uncertainty as a flux per steradian (intensity) rather than flux per beam area.  Evaluating the beam solid angle using

\begin{equation}
    \label{eq:beam area}
     \Omega = \left(\frac{\pi}{4\ln{2}}\right)(FWHM)^2 \approx 1.13(FWHM)^2
\end{equation}

 we obtain

\begin{equation}
    \label{eq:mapping_speed_total_intensity_approx_2}
     \sigma_{I} = \left(NEFD_{phot}\right) \left(\frac{(\pi/180) }{(3600 s)^{0.5}(1.13)^{1.5}}\right) (FWHM/rad)^{-3} (N_{det}Yield)^{-0.5}.
\end{equation}

The corresponding quantity for polarized flux is given by

\begin{equation}
    \label{eq:mapping_speed_polarized_intensity_approx}
     \sigma_{P} = \frac{\sqrt{2}\sigma_{I}}{\eta_P}
\end{equation}

where the factor $\sqrt{2}$ is due to the same reason discussed previously in \ref{eq:sigma_P}.

\begin{figure}[t]
    \centering
    \subfigure[$175\mu m$]{
          \includegraphics[scale=0.5]{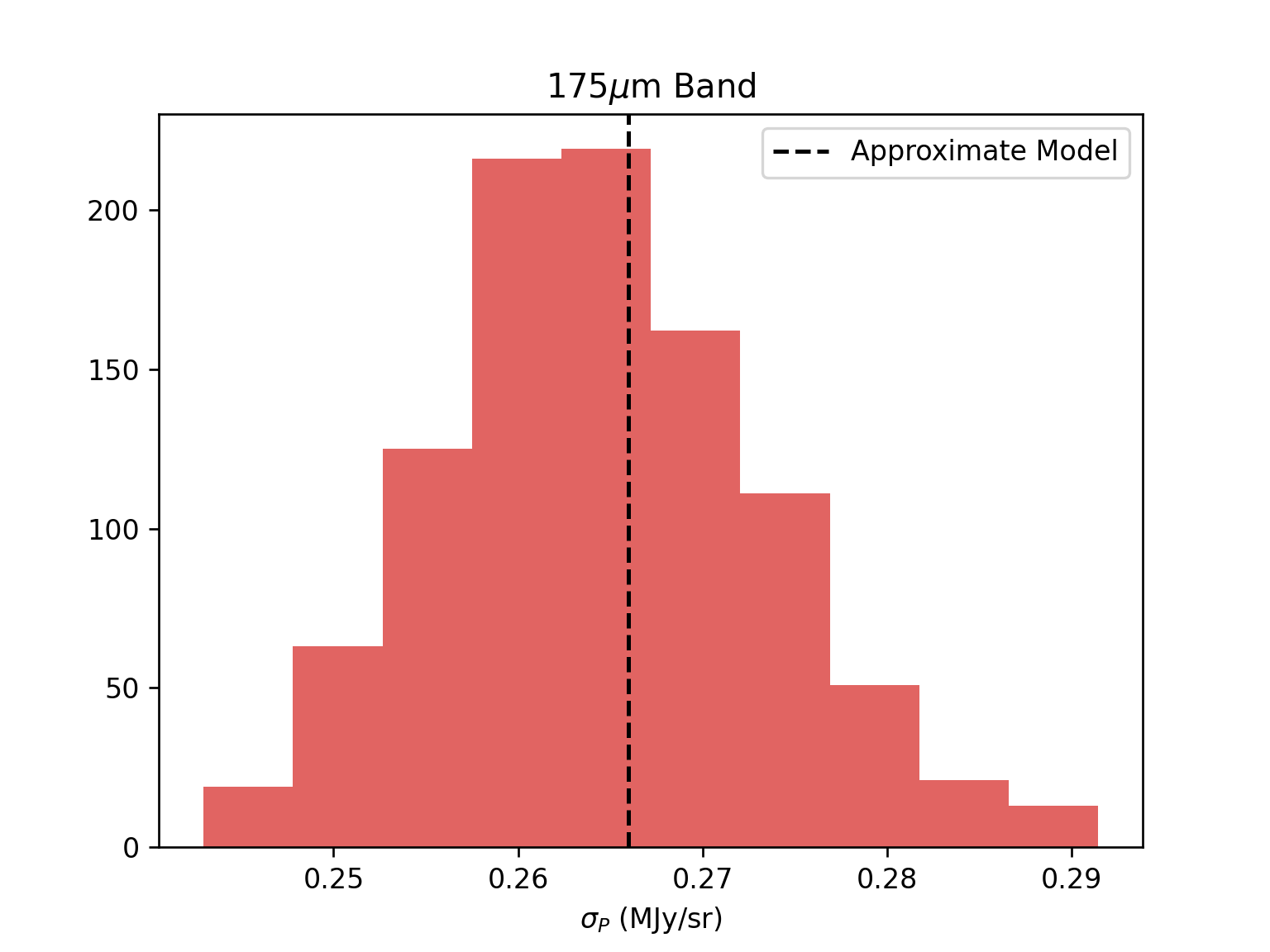}
          \label{fig:sigma_175}}
    \hfill
    \subfigure[$250\mu m$]{
          \includegraphics[scale=0.5]{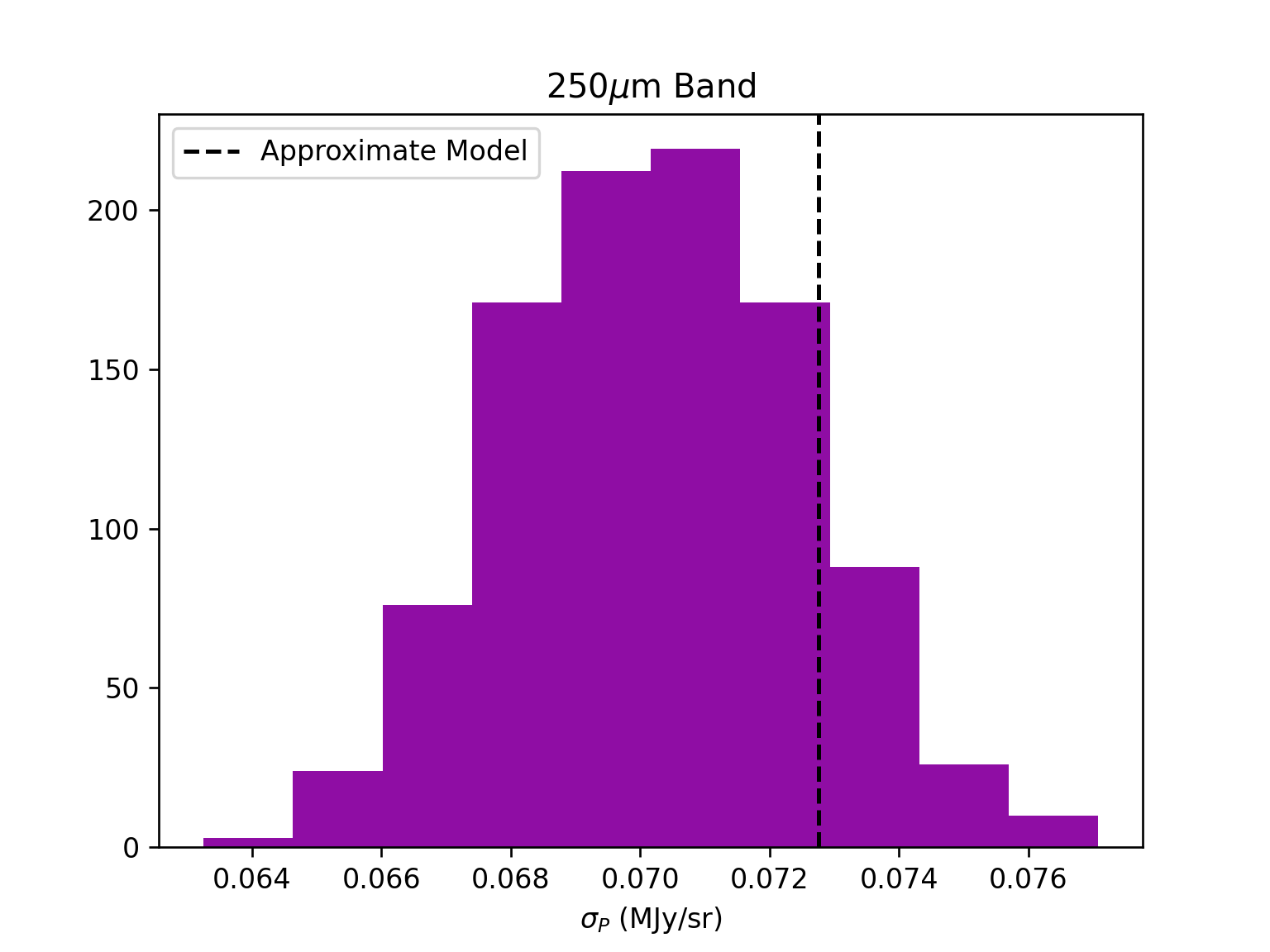}
          \label{fig:sigma_250}
    }\\
    \subfigure[$350\mu m$]{
          \includegraphics[scale=0.5]{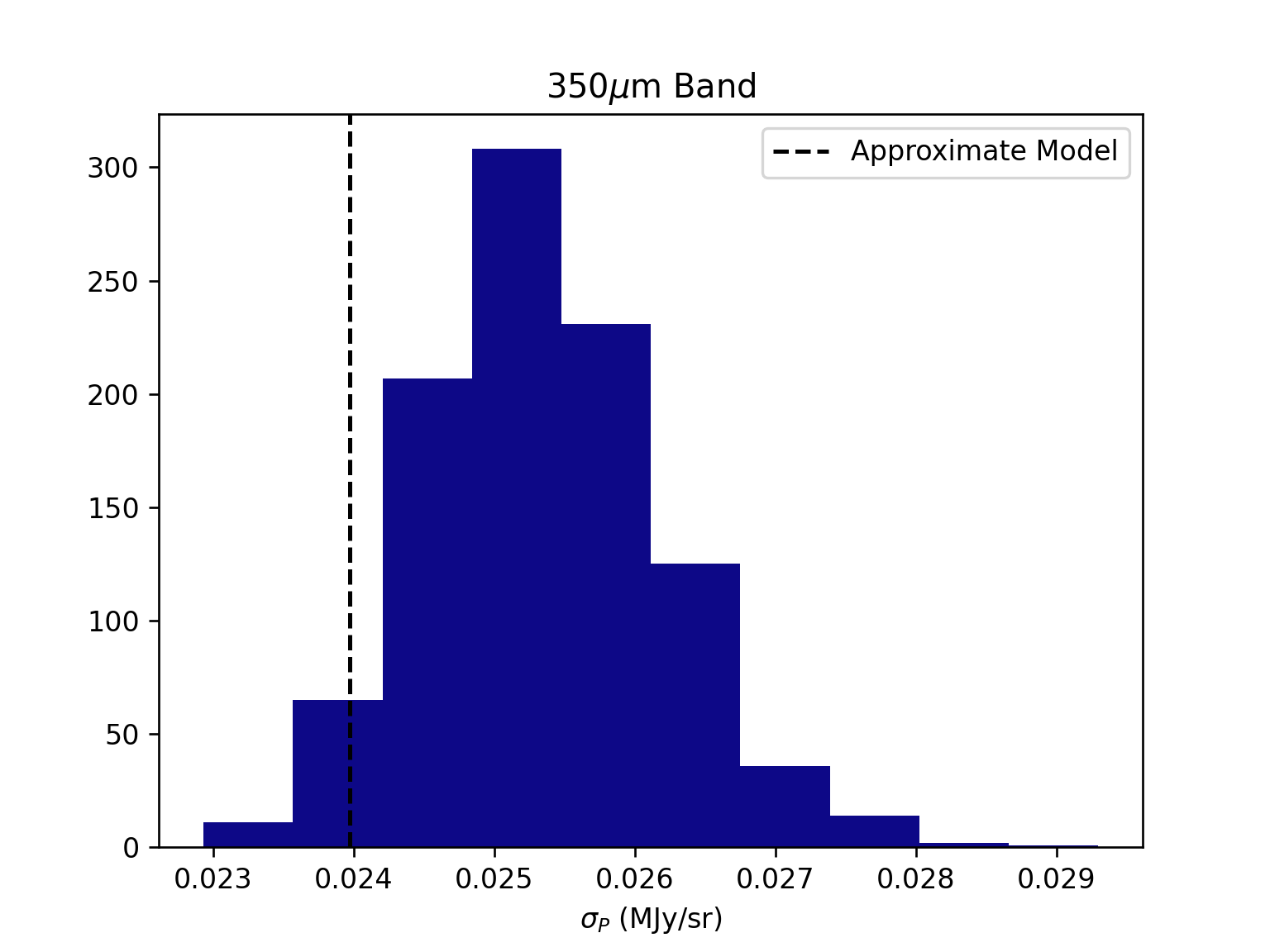}
          \label{fig:sigma_350}
    }
    \caption{Histograms from the Monte Carlo simulation of the $\sigma_P$ at each band. The result of the approximate method is also presented as a vertical line.}
    \label{fig:resultsMC}
\end{figure}

\begin{table}[t]
\vspace{-0.2in}
     \scriptsize
    \centering
    \begin{tabular}{lcccccc}
        \toprule
        \textbf{FoM} & \multicolumn{3}{c}{\textbf{Analytical Model}} &  \multicolumn{3}{c}{\textbf{Approximate Model}}\\
         & 175$\mu$m & 250$\mu$m & 350$\mu$m & 175$\mu$m & 250$\mu$m & 350$\mu$m \\
        \midrule
        $NEP (10^{-18} W/\sqrt{HZ})$ & $5.06^{+0.22}_{-0.21}$ & $4.14^{+0.17}_{-0.17}$ & $4.05^{+0.17}_{-0.17}$ & $4.42$ & $3.79$ & $3.41$\\
        $NEFD (mJy\cdot\sqrt{s})$ & $1.652^{+0.073}_{-0.068}$ & $1.279^{+0.053}_{-0.053}$ & $1.266^{+0.055}_{-0.051}$ & $1.661$ & $1.324$ & $1.197$ \\
        $\sigma_I (MJy/sr)$ & $0.1683^{+0.0074}_{-0.0069}$ & $0.0447^{+0.0018}_{-0.0018}$ & $0.0161^{+0.0007}_{-0.0006}$ & $0.1692$ & $0.0463$ & $0.0152$\\
        $\sigma_P (MJy/sr)$ & $0.26454^{+0.0117}_{-0.0108}$ & $0.0703^{+0.0029}_{-0.0029}$ & $0.0253^{+0.0011}_{-0.0011}$ & $0.2659$ & $0.0727$ & $0.0239$\\
        \bottomrule
    \end{tabular}
    \caption{Results from the sensitivity models. Both the NEP and NEFD refers to the array and in intensity.}
    \label{tab:results}
\end{table}

\subsection{Results}

We estimate the performance of the instrument in the three BLAST Observatory frequency bands using both models. We considered an elevation of 40$^\circ$, which is the expected average for BLAST Observatory, for the calculations of the effect of the atmosphere on both models, the analytical and the approximate. In this condition, the atmosphere has an average emissivity of $2\%$, $1.5\%$ and $0.6\%$ at 175$\mu$m, 250$\mu$m and 350$\mu$m, respectively. For the approximate models, the assumptions and resultant values are presented in the Table \ref{tab:sanity_check}. For the analytical model, we performed a Monte Carlo simulation where we varied randomly the yield of the array and the optical efficiency of each element in the instrument. For the detector yield, we considered a Gaussian distribution centered at $80\%$ with a standard deviation of $5\%$. For the filters and the other optical elements that has a frequency dependent curve of the optical efficiency, we introduced for each frequency point in the curve we introduced a perturbation. This one is assumed to be gaussian, centered on the measured optical efficiency and a standard deviation of $2\%$. A comprehensive list of the optical elements included in the analytical model calculation is presented in Table \ref{tab:optical_elements} and the final instrumental optical efficiency as estimated by the analytical model is presented in Figure \ref{fig:opticalefficiency}. We also considered an imperfect primary and secondary mirror with a surface roughness of $7\mu$m and $5\mu$m, respectively.

The figure of merits for the instrument from both models are presented in Table \ref{tab:results}. The values for the $\sigma_{I/P}$ are calculated for an area of $1\,deg^2$ and $1\,h$ of integration time. The results from the two models are in general compatible, except for the NEP. The differences can be explained looking at Fig.\ref{fig:opticalefficiency}. Indeed, it is possible to notice that the optical efficiency used by the approximate model underestimates the instrumental bandpass. For the $\sigma_P$ figure of merit, we report also the histograms from the Monte Carlo simulations in Fig.\ref{fig:resultsMC}. 
Finally, Fig.\ref{fig:comparisonplot} compares our BLAST Observatory sensitivity forecasts with sensitivity values for other recent, current, and planned facilities.  As noted in the figure caption, the sensitivity information for other facilities was taken from the respective forecast publications and/or on-line sensitivity calculators.  

\input{Figure1}

%% file: Figure1.tex
\begin{figure}[t]
\begin{center}
%\begin{minipage}[c]{0.6\textwidth}
 \centering
 \includegraphics[width=0.8\textwidth,trim=0.25in 0.25in 0.25in 0.5in]{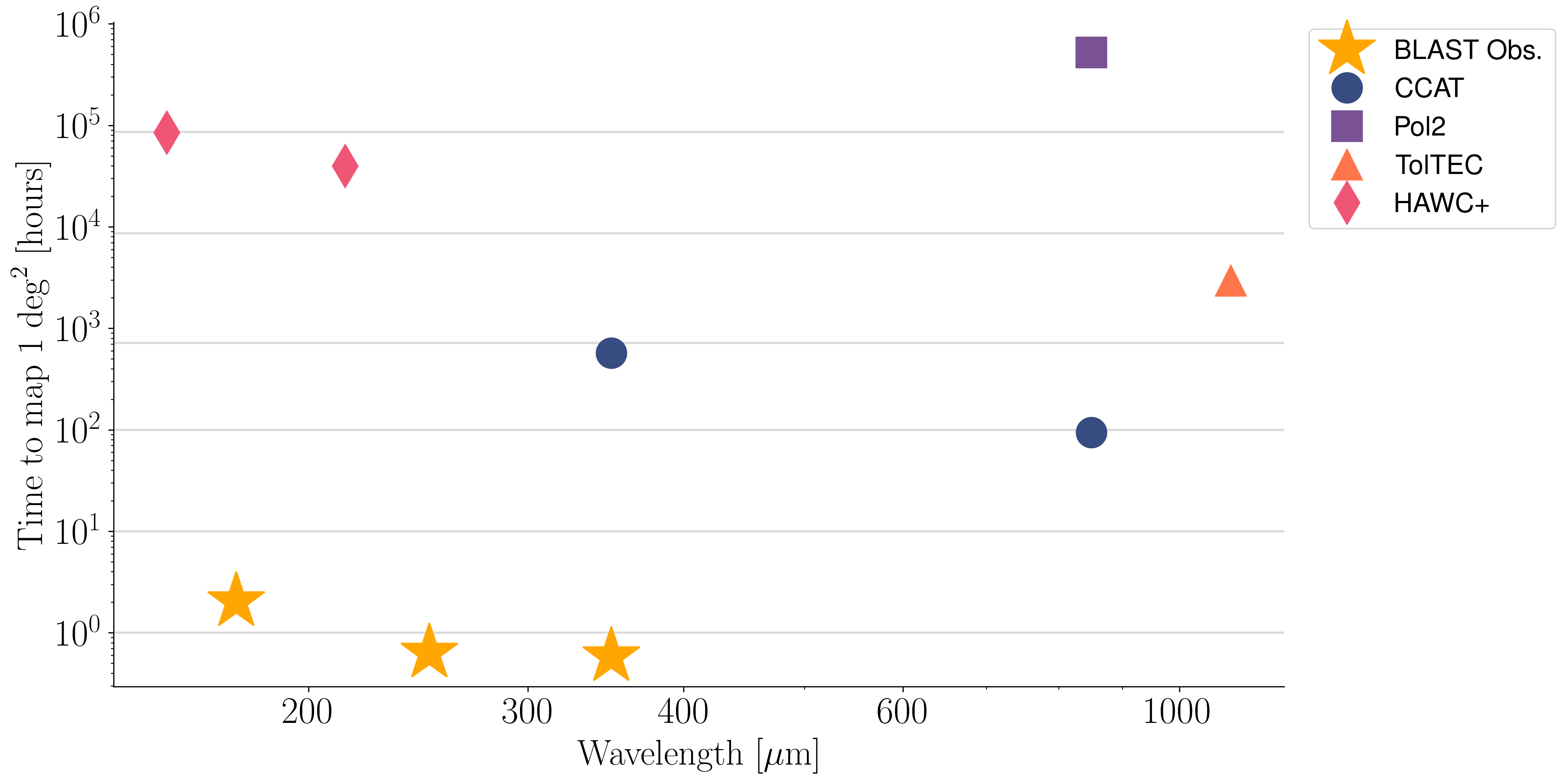}
 %{Figures/comparison_plot_wavelength_1sqdg_2022update_revised_grayscale_noBPol_JWSTgap_extrawords.pdf}
%\end{minipage}\hfill

%\begin{minipage}[c]{0.37\textwidth}
\vspace{0.2cm}
    \caption {The BLAST Observatory provides a vast improvement in mapping speed over current and planned facilities by exploiting the ultra-low backgrounds and wide spectral bands achievable from a balloon platform. Vertical axis shows estimated time to detect dust polarization at $3\sigma$ for a 1 deg$^2$ region of diffuse ISM, assuming the SED in \cite{HensleyDraine:2021} scaled to a 350~$\mu$m polarized intensity of 0.1 MJy\,sr$^{-1}$. All estimates assume smoothing to the BLAST Obs.~350 $\mu$m resolution of $55''$. \\
    BLAST Observatory sensitivity based on calculations described in Section \ref{sec:sensitivity_model}; CCAT-prime estimates are based on \cite{PrimeCam2023} who assume median atmosphere for the best 3 quartiles (350\,$\mu$m band transmission $\sim$25\%); POL-2, TolTEC, and HAWC+/SOFIA estimates are based on SCUBA-2 Integration Time Calculator, TolTEC maximum mapping speed (U.\ Mass.\ web pages), and HAWC+ MIfP from \cite{harper2018}, respectively.
    }

    %\caption {\small{\textbf{The BLAST Observatory achieves an orders-of-magnitude improvement in mapping speed above both the current state of the art and future ground-based facilities.} Comparison to current and future polarimeters: Estimated time to detect dust polarization at $3\sigma$ for a 1 deg$^2$ region of the diffuse, high-Galactic latitude ISM, assuming the data-constrained SED in \citet{HensleyDraine:2021}. Sensitivities are estimated for maps smoothed to the BLAST Obs.~350 $\mu$m resolution of FWHM=$55''$. }\\
    %The BLAST Observatory is the \textit{only} facility that can map regions of diffuse polarized dust emission with high angular resolution, enabling unique measurements of dust composition (SO1) and interstellar turbulence (SO2). BLAST Obs. will also make deep, complete maps of entire molecular clouds (SO3). Toward dense, cold regions where it is feasible, other facilities are synergistic with BLAST Observatory because they can map small regions at higher angular resolution (FWHM=14$''$ with POL-2 and $5''$ with TolTEC). With the cancellation of SOFIA, BLAST Obs. is the \textit{only} instrument in this crucial wavelength regime.} \\
    %\footnotesize{CCAT-prime estimates are based on Ref. \cite{choi20} who assume median atmosphere for the best 3 quartiles (350\,$\mu$m band transmission $\sim$25\%).
    %POL-2, TolTEC, and SOFIA estimates are based on SCUBA-2 Integration Time Calculator and maximum mapping speed prediction on TolTEC (U.\ Mass.\ web pages, and SOFIA Cycle 10 call for proposals, respectively).
    %}
    %}
    \label{fig:comparisonplot}
%\end{minipage}
\end{center}

\end{figure}

%% file: section_conclusions.tex
\section{Conclusions}
\label{sec:conclusions}

We have presented the design and high level scientific goals of a proposed balloon-borne mission, BLAST Observatory. The sensitivity models and methodologies presented here can be used to make forecasts of scientific returns and the potential of the instrument for discoveries in this unique observational space. We leave to future publications more detailed forecasting results for each of the principal science cases.

%% file: acknowledgments.tex
\begin{acknowledgments}

Part of this work was funded by NASA Grant 21-APRA21-0020 Cooperative Agreement Number 80NSSC22K1845. %, The BLAST (Balloon-borne Large Aperture Submillimeter Telescope) Observatory, Cooperative Agreement Number 80NSSC22K1845 P00002
GC acknowledges the support given by the European Research Council under the Marie Sklodowska Curie actions through the Individual European Fellowship No. 892174 PROTOCALC for part of this work. 

L.M.F. acknowledges support from an NSERC Discovery Grant (RGPIN-2020-06266) as well as a Research Initiation Grant from Queen’s University.

This work was carried out in part at the Jet Propulsion Laboratory, California Institute of Technology, under a contract with the National Aeronautics and Space Administration.

\end{acknowledgments}

%% file: appendix_sensitivity_assumptions.tex
\section{Assumptions for the Sensitivity Model}\label{sec:ass}

\subsection{Emissivity Estimation}\label{subsec:absorption}

To estimate the absorption of the different optical elements, the model assumes that the output power is:

\begin{equation}
    \label{eq:output}
    I_{out} = I_0\cdot\left(1-\rho\right)e^{-\alpha t}
\end{equation}

where $I_0$ is the emitted power, $\rho$ is the reflectivity, $\alpha$ is the absorption coefficent and $t$ is the thickness of the optical element. This gives a transmissivity:

\begin{equation}
    \label{eq:trans_c}
    \tau = \left(1-\rho\right)e^{-\alpha t}
\end{equation}

The value of $\tau$ has been measured by Cardiff for a specific thickness and this allow to estimate the reflectivity which is in general indipendent of the thickness. 
This reflectivity is used to compute the transmissivity at different thicknesses as:

\begin{equation}
    \label{eq:trans}
    \tau = \tau_{c}e^{\alpha \left(t_c-t\right)}
\end{equation}

where the subscript $c$ refers to the thickness and transmissivity values from the Cardiff sample. 
Using eq.\ref{eq:trans} it is possible to estimate the emissivity as:

\begin{equation}
    \label{eq:emiss}
    \epsilon = 1-\tau-\rho
\end{equation}

\subsection{Scattering on the surfaces}\label{subsec:scatter}

The scattering coefficient of a mirror can be estimated using the Ruze formula:

\begin{equation}
    \label{eq:ruze}
    S = e^{-\left(\frac{4\pi\sigma}{\lambda}\right)^2}
\end{equation}

where $\sigma$ is the RMS of the surface.